\def\conference{0}
\def\shownotes{0}

\ifnum\conference=0
\documentclass[sigconf,authorversion]{acmart}
\else
\documentclass[sigconf]{acmart}
\fi
	
	\newtheorem*{theorem*}{Theorem}

\newcommand{\myparab}[1]{\vspace{1ex}\noindent\textbf{#1.}}

\usepackage{multirow}
\usepackage{algorithm}
\usepackage{algorithmic}
\usepackage{graphicx}
\usepackage{textcomp}
\usepackage{xcolor,colortbl}
\usepackage{subcaption}
\usepackage{hyperref}
\usepackage{xspace}
\usepackage{float}
\usepackage{url}

\hypersetup{
    colorlinks=true,
    linkcolor=blue,
    filecolor=magenta,      
    urlcolor=cyan,
    pdftitle={Adversarial Paper},
    pdfpagemode=FullScreen,
    }

\definecolor{Grayl}{gray}{0.65}
\definecolor{Gray}{gray}{0.85}
\newcolumntype{a}{>{\columncolor{Gray}}r}
\newcolumntype{b}{>{\columncolor{Grayl}}r}

\def\BibTeX{{\rm B\kern-.05em{\sc i\kern-.025em b}\kern-.08em
    T\kern-.1667em\lower.7ex\hbox{E}\kern-.125emX}}

\newcommand{\combinedRGB}{\textit{RGB-C}\xspace}
\newcommand{\bubblesRGB}{\textit{RGB-B}\xspace}
\newcommand{\combinedGray}{\textit{Gray-C}\xspace}
\newcommand{\bubblesGray}{\textit{Gray-B}\xspace}
\newcommand{\advmark}{\textit{Over}\xspace}
\newcommand{\advnmark}{\textit{Under}\xspace}

\newcommand{\combined}{\textit{Combined}\xspace}
\newcommand{\bubbles}{\textit{Bubbles}\xspace}

\newcommand{\apgd}{\ensuremath{\mathtt{APGD}}\xspace}

\newcommand{\pgd}{\ensuremath{\mathtt{PGD}}\xspace}
\newcommand{\fgsm}{\ensuremath{\mathtt{FGSM}}\xspace}
\newcommand{\mim}{\ensuremath{\mathtt{MIM}}\xspace}

\ifnum\shownotes=0
\newcommand{\authnote}[2]{{\marginpar{\tiny \textcolor{red}{\textsf{#1 notes: }\textcolor{blue}{#2}}}}}
\else
\newcommand{\authnote}[2]{}
\fi

\begin{abstract}
We show the security risk associated with using machine learning classifiers in United States election tabulators. The central classification task in election tabulation is deciding whether a \emph{mark} does or does not appear on a \emph{bubble} associated to an alternative in a contest on the ballot. Barretto et al. (E-Vote-ID 2021) reported that convolutional neural networks are a viable option in this field, as they outperform simple feature-based classifiers. 

Our contributions to election security can be divided into four parts. To demonstrate and analyze the hypothetical vulnerability of machine learning models on election tabulators, we first introduce four new ballot datasets. Second, we train and test a variety of different models on our new datasets. These models include support vector machines, convolutional neural networks (a basic CNN, VGG and ResNet), and vision transformers (Twins and CaiT). Third, using our new datasets and trained models, we demonstrate that traditional white box attacks are ineffective in the voting domain due to gradient masking. Our analyses further reveal that gradient masking is a product of numerical instability. We use a modified difference of logits ratio loss to overcome this issue (Croce and Hein, ICML 2020). Fourth, in the physical world, we conduct attacks with the adversarial examples generated using our new methods. In traditional adversarial machine learning, a high (50\% or greater) attack success rate is ideal. However, for certain elections, even a 5\% attack success rate can flip the outcome of a race. We show such an impact is possible in the physical domain. We thoroughly discuss attack realism, and the challenges and practicality associated with printing and scanning ballot adversarial examples.
\end{abstract}
\author{Kaleel Mahmood}
\orcid{0000-0002-7672-4449}
\affiliation{\institution{University of Rhode Island}
\department{Department of Computer Science and Statistics}
\city{Kingston}
\state{Rhode Island}
\country{United States}}
\email{kaleel.mahmood@uri.edu}

\author{Caleb Manicke}
\orcid{0009-0007-6409-5343}
\affiliation{\institution{University of Connecticut}
\department{Voting Technology Center}
\city{Storrs}
\state{Connecticut}
\country{United States}}
\email{caleb.manicke@uconn.edu}

\author{Ethan Rathbun}
\affiliation{\institution{Northeastern University}
\orcid{0000-0002-5437-2489}
\department{Khoury College of Computer Sciences}
\city{Boston}
\state{Massachusetts}
\country{United States}}
\email{rathbun.e@northeastern.edu}

\author{Aayushi Verma}
\orcid{0000-0003-2396-4569}
\affiliation{\institution{University of Connecticut}
\department{Voting Technology Center}
\city{Storrs}
\state{Connecticut}
\country{United States}}
\email{aayushi.verma@uconn.edu}

\author{Sohaib Ahmad}
\orcid{0000-0001-6720-0507}
\affiliation{\institution{University of Connecticut}
\department{Voting Technology Center}
\city{Storrs}
\state{Connecticut}
\country{United States}}
\email{sohaib.ahmad@uconn.edu}

\author{Nicholas Stamatakis}
\orcid{0009-0000-1440-7192}
\affiliation{\institution{Stony Brook University}
\department{Department of Computer Science}
\city{Stony Brook}
\state{New York}
\country{United States}}
\email{nikola268345@gmail.com}

\author{Laurent Michel}
\orcid{0000-0001-7230-7130}
\affiliation{Synchrony Chair in Cybersecurity
\institution{University of Connecticut}
\department{Voting Technology Center}
\city{Storrs}
\state{Connecticut}
\country{United States}}
\email{laurent.michel@uconn.edu}

\author{Benjamin Fuller}
\orcid{0000-0001-6450-0088}
\affiliation{\institution{University of Connecticut}
\department{Voting Technology Center}
\city{Storrs}
\state{Connecticut}
\country{United States}}
\email{benjamin.fuller@uconn.edu}

\copyrightyear{2025}
\acmYear{2025}
\setcopyright{rightsretained}
\acmConference[CCS '25] {Proceedings of the 2025 ACM SIGSAC Conference on Computer and Communications Security}{ October 13--17, 2025}{Taipei, Taiwan.}
\acmBooktitle{Proceedings of the 2025 ACM SIGSAC Conference on Computer and Communications Security (CCS '25), October 13--17, 2025, Taipei, Taiwan}

\acmDOI{10.1145/3719027.3744882}
\acmISBN{979-8-4007-1525-9/2025/10}
\ifnum\conference=1

\fi

\begin{document}

\title[Busting the Paper Ballot]{Busting the Paper Ballot: Voting Meets Adversarial Machine Learning}
\renewcommand{\shorttitle}{Busting the Paper Ballot}
\renewcommand{\shortauthors}{Mahmood, Manicke, Rathbun, Verma, Ahmad, Stamatakis, Michel, Fuller}

\maketitle

\section{Introduction}
Elections in the United States are decentralized and conducted by the states. 
The \href{https://www.eac.gov/about/help\_america\_vote\_act.aspx}{Help America Vote Act} of 2002 prompted all states to modernize their voting infrastructure and retire lever machines. States overwhelmingly adopted voter marked paper ballots that yield a ``voter verifiable paper audit tail'' or VVPAT~\cite{rivest2017election} and are scanned and counted by digital tabulators.  
According to \href{https://verifiedvoting.org/verifier/#mode/navigate/map/ppEquip/mapType/normal/year/2024}{Verified Voting Database}, 69.2\% of tabulators are scanners and 25.9\% are ballot marking devices (BMD), leaving only a 4.9\% market share for direct recording devices that do not use paper at all.\footnote{We exclude ballot marking devices that print two versions of the voter's preferences: one readable by the voter and a machine interpretation such as a QR code that is ingested by the tabulation.}
To assess the voter selection for any contest on a ballot, the tabulator determines whether \emph{bubbles} associated to alternatives in the contest are \emph{blank} or \emph{marked}. Namely, the core task is to carry out a binary classification on the digital image of a bubble. 
Barreto et al. argued that Convolutional Neural Networks (CNNs) ~\cite{CNNBallot} are suitable for ballot mark recognition with up to $99.9\%$ accuracy on manually labeled ballots. However, machine learning classifiers are vulnerable to adversarial examples~\cite{goodfellow2014explaining} where an imperceptible perturbation added to the input induces a misclassification.

\emph{This paper explores the susceptibility of machine learning classifiers to adversarial attacks when presented with images of bubbles from the voting domain.} 
Specifically, our paper introduces attacks where one can implant adversarial machine learning examples~\cite{szegedy2013intriguing,goodfellow2014explaining,mcdaniel2016machine} on ballots handled by the voter before it is fed to the tabulator. An adversarial signal is visually imperceptible (to a human),
yet alters the classification results. \emph{We focus on attacks that could be conducted by a compromised vendor that prints ballots.}  The attacker's goal is to print a ballot that appears empty but where some bubbles are interpreted as marked by the tabulator.  We detail our threat model in Section~\ref{sec:adv attacks}.
This voting domain is unique for the following reasons: 
\begin{enumerate}
    \itemsep0em
    \item It focuses on a deceptively simple binary classification.
     \item \href{https://www.eac.gov/voting-equipment/voluntary-voting-system-guidelines}{Voluntary voting system guidelines (VVSG) 2.0} require publication by vendors of the mechanisms used to classify a bubble,\footnote{Requirement 1.1.6G and accompanying discussion.} making \emph{white box attacks} realistic.
     \item The attacker has to print a {\it signal} on paper which is scanned. Effects such as printer dithering must be considered. Kurakin et al.~\cite{kurakin2018adversarial} previously considered attacks in the physical world. Our physical world setting differs from prior work.
    \item There are no agreed-upon labeled datasets for mark classification.  Human auditors are expected to capture voter intent with guidelines that vary by state (see discussion in Section~\ref{sec:bubbles}). 
    \item  An attacker can freely reuse adversarial examples; bubbles printed on a ballot are supposed to be identical.
    \item Election equipment has a long life cycle. Deploying vulnerable models carries lasting risks.
\end{enumerate}

We conduct our attacks on six representative models, a support vector machine (SVM), a three-layer CNN that we call SimpleCNN, VGG-16 CNN~\cite{VGG16}, a ResNet-20 CNN~\cite{RESNET}, Class Attention in Image Transformer (CaiT)~\cite{touvron2021going}, and the Twins vision transformer~\cite{twins}.

\myparab{Our Contribution} We demonstrate the hypothetical vulnerability of using machine learning classifiers in bubble recognition in both the digital and physical setting. In doing so we make the following contributions:
\begin{enumerate}
\item \textbf{New Labeled Voting Datasets:} We introduce four new labeled ballot datasets (two grayscale and two color) for training machine learning classifiers for ballot mark recognition~\cite{uconn_voter_center_2025_15458710}. 

\item \textbf{Gradient Masking on Voting Datasets:} We show that for the three convolutional models, the conventional application of white-box attacks (APGD~\cite{APGD}, PGD~\cite{madry2018towards} and MIM~\cite{MIM}) does not work. Models show robustness $>0$ when the adversary can apply unbounded perturbations. This failure is 
attributed to numerical instability causing gradients in backpropagation to be reported as $0$, despite the models achieving high accuracy during training. To the best of our knowledge, all previous examples of gradient masking occurred via defensive methods to stop adversarial examples~\cite{carlini2017adversarial}.

\item \textbf{Overcoming Gradient Masking:} We modify the difference of logits ratio (DLR) proposed by Croce and Hein~\cite{APGD} to work for binary classification (our modification can be viewed as an untargeted version of Carlini and Wagner's loss~\cite{carlini2017towards}).

\item \textbf{Physical Attacks:} We show that the printing and scanning process (using commodity equipment) drastically degrades the adversarial attack signal.  Despite this, some attacks on some models are effective enough to still impact election races with small margins where many voters do not specify a preference.
\end{enumerate}

\myparab{Disclaimer} We intentionally target common classification models rather than any model used by a tabulator.  No tabulator manufacturers has been certified to \href{https://www.eac.gov/voting-equipment/voluntary-voting-system-guidelines}{VVSG 2.0}, so details are not yet available on their classification methods. The purpose of this work is to highlight the risk in potentially deploying machine learning models in these systems. As discussed in Appendix~\ref{sec:ethics} on ethical considerations, our target machine learning models are chosen to cover the design space.  They are not an attempt to recreate choices made by vendors. Ballot printers are specialized vendors with long-term relationships with municipalities. We believe an external attack on a vendor is more likely than an insider intentionally compromising ballot printing. 

\myparab{Paper Organization} Section~\ref{sec:voting overview} details election systems in the United States. Section~\ref{sec:dataset_classifiers} present the new datasets and classifiers under test. Section~\ref{sec:adv attacks} describes our adversarial threat model and adversarial example generation methods. Section~\ref{sec:gradient masking evidence} shows that gradient masking occurs on our datasets after standard training. Section~\ref{sec:gradient masking} how DLR overcomes gradient masking. Section~\ref{sec:digital domain} presents our experimental results and analyses in the digital domain. Section~\ref{sec:physical domain} details our physical domain attack results. Section~\ref{sec:conclusion} concludes and presents open questions. Our Appendix contains further experimental details. Classifiers under test and attacks are at \url{https://github.com/VoterCenter/Busting-the-Ballot}.

\section{Voting in the United States and Prior Work}
\label{sec:voting overview}
This section provides an overview of voting practices and related security research in the United States. Voting processes are meant to enforce the 1-voter to 1-vote principle to assure fairness. Voting by mail and voting in person rely on different processes suitable to each modality. This work focuses on in-person voting. This process 
involves multiple steps 1) checking-in voters, 2) handing out ballots, 3) voting and casting of a  ballot and 4) tabulating the results. This paper further restricts its scope to the latter steps of this pipeline: voting, casting and tabulating.
Voting systems used in the U.S. include:\footnote{See an overview at \href{https://verifiedvoting.org/votingequipment/}{Verified Voting}.  
Observe that smaller categories exist, including DREs, that create voter-verifiable paper trails.}\footnote{Eliding research on Internet~\cite{specter2020ballot} and cryptographic~\cite{benaloh1987verifiable,adida2008helios,benaloh2024electionguard} voting.}

\myparab{Ballot Scanners} Take in ballots with bubbles and determine which bubbles on the ballot have been filled in. The appeal of scanners (used in  $66.6\%$ of the U.S.)  is that ballots are typically marked directly by voters and form  a VVPAT.  The VVPAT can be used for machine independent audits.

\myparab{Ballot Marking Devices} Provide the necessary means to \textit{fill} bubbles on behalf of the voter based on an alternate (often digital and computerized) input mechanism instead of a pen. A voter using a BMD fills in a digital artifact, that creates a printout of a filled ballot. Some BMDs fill in bubbles and rely on conventional paper ballots. Others encode the ballot in a machine readable format (like a QR code, or a barcode) or print the selection in each race as plain text.\footnote{A recent executive order, \href{https://www.whitehouse.gov/presidential-actions/2025/03/preserving-and-protecting-the-integrity-of-american-elections/}{Preserving and Protecting the Integrity of American Elections}, asks for all preferences to be printed in plain text.} 
Selections, conveyed through filled bubbles, are interpreted primarily by a tabulator as in the previous method. Precincts with BMDs account for $25.9\%$ of the U.S. electorate.  
    
\myparab{Direct Recording Equipment} Voters use an interface to state their preferences and the machine records these preferences and adds them to a tabulation. The DRE device is used for all 3 stages: encoding a ballot, casting the ballot and tabulating all the ballots.
There is no paper artifact of the voter's preferences. 
These systems are used in roughly $4\%$ of the U.S.\footnote{There are variants not reviewed here so the numbers do not add to $100\%$.} Our work does not apply to DREs, but these systems are shunned due to the lack of a VVPAT~\cite{national2018securing}.

\subsection{Types of Ballot Scanners}
\label{ssec:ballot scanners}
\noindent Ballot scanners are given a page containing
several {\it questions}, each with multiple outcomes that can be chosen by filling a bubble. The scanner classifies each bubble as a blank or a marked bubble. These determinations are then used with election-specific rules to determine the votes. 

\myparab{Optical Lens Systems} This analog technology is ubiquitous in standardized tests where examinees fill bubble sheets that convey answers. Each page features {\it timing marks} (black rectangles) on its edges to hint at the position of logical rows and columns. All the bubbles on the page form a matrix and are addressable based on their row and column indices. Scanners process one row at a time (when facing a row timing mark), effectively opening an analog sensor to collect reflected light. The machine has a light sensor in each column position. If a bubble is filled, it absorbs more light than a blank bubble.
Whether bubbles are read as marked depends on the thresholds and sensitivity of the sensors. Image segmentation is a function of timing marks and sensor position.

\myparab{Full Image Scanners} Off-the-shelves full-image scanners are also used.
Modern hardware can record anything from 100 dots-per-inch to 1200 dots-per-inch. Given a 8x11 US letter page, a 200 DPI resolution implies that each row has $1700=8.5\cdot200$ pixels and a grand total of $2200=11\cdot 200$ rows.
Sensors typically captures several rows at a time and the pixels are physically arranged in a so called {\it Bayer} pattern. The firmware of the scanner (or the driver on the host computer) converts the Bayer matrix of grayscale values into a matrix of RGB values by reconstructing the missing color values through interpolation. This process is known as debayering or demosaicing~\cite{kwan2017debayering}. The end result is a RGB pixels where each color channel is grayscale and uses 8 bits per pixel. Pulling the page over the  sensor is a mechanical process subject to acceleration and deceleration. The sensor sensitivity impacts the color rendition of the device.
\label{ssec:image:analysis}
Once acquired by a COTS Scanner, an image is analyzed using the following steps:
\begin{itemize}
\itemsep0em
\item \emph{Stretching} The raw image has an effective DPI rate that varies with the acceleration of the ballot. A correction {\it inverts} this stretch to bring features closer to their true relative location. 
\item \emph{Registration} A constellation of geometric features identified on a reference ballot are used to align any incoming scan with the reference image. Once this is done, a bubble at coordinate $(x,y)$ in the reference image is expected to be at coordinate $(x,y)$ in the registered scan. 
\item \emph{Chromatic Correction} An \href{https://en.wikipedia.org/wiki/ICC_profile}{ICC profile} corrects the colors in the image to bring them as close as possible to the true colors based on a device specific colorimetry calibration.
\item \emph{Segmentation} The location of the bubbles on the reference image is used to locate and clip out small RGB bitmaps that contain the actual bubbles. 
\item \emph{Classification} The final step classifies each bitmap as a blank bubble or a marked bubble.
\end{itemize}
Once bubbles are classified as blanks or marks, each race on the ballot can be tabulated according to the rules of the race. 
The ultimate tabulation stage is not the object of this paper. We also assume that all stages up to and including segmentation are accurate. 

\myparab{Prior Work}
Tabulation security received renewed attention~\cite{kohno2004analysis,hursti2005black,wagner2006security,antonyan2009state,jancewicz2013malicious,davtyan2008pre,davtyan2009taking,checkoway2009can,antonyan2010determining,buell2011auditing} after the \href{https://www.eac.gov/about/help_america_vote_act.aspx}{Help America Vote Act} in 2002. Issues ranged from unprotected serial ports, manipulation of election definitions, and exploitation of poorly designed cryptography.  Procedures including risk-limiting audits or RLAs~\cite{lindeman2012gentle,zagorski2021minerva,broadrick2022simulations,harrison2022adaptive,fuller2024decisive} were created to deal with these vulnerabilities. Note that RLAs only detect \emph{whether there is an error in the reported outcome}.  Detecting the root cause of such errors can be complicated or impossible.  Other classes of vulnerabilities include lax adherence to policy~\cite{halderman2021analysis}.  Procedures and requirements are formalized in voluntary voting system guidelines or VVSG~\cite{skall2005voluntary,tatum2018abcs,quesenbery2023handbook}.  

Vulnerabilities persists in modern systems~\cite{bernhard2017public} including the fact that voters do not always inspect the output of BMDs~\cite{bernhard2019unclearballot,bernhard2020can}.  Imprinting of identifiers on ballots at tabulation time, which enables more efficient RLAs (see discussion in \cite{harrison2022adaptive}), also requires careful use of cryptography~\cite{crimmins2024dvsorder}.  These vulnerabilities have rightly placed the design of secure ballot tabulation devices as a primary focus for the community.  

\section{Datasets / Classifier Architectures}
\label{sec:dataset_classifiers}
This section introduces four new datasets of segmented regions on a ballot for interpretation. These are images of \emph{bubbles}. For each dataset, we define its properties and utility for further investigation. 
Datasets and the accompanying software are released alongside this paper.
The remainder of the section reviews each ML classifiers and explains the purpose for inclusion in the analyses. 

\subsection{New Bubble Datasets}
\label{sec:bubbles}
Images are segmented using ballot geometry. We do not consider image segmentation as an attack target.
Mark interpretation is a supervised binary classification problem, requiring representative datasets of marks and empty bubbles~\cite{CNNBallot}. 
The segmented images are $40 \times 50$ pixels. 
Fully darkened bubbles should be interpreted as marks while empty bubbles should be interpreted as nonmarks. Naturally, one would need some separation oracle to define a boundary between marks and nonmarks. No matter what oracle is chosen, some images will be very \textit{close} to the boundary. Such images are called \emph{marginal marks}~\cite{bajcsy2015systematic} and may include 
samples such as checkmarks, crosses, lightly filled or even accidentally filled bubbles, see~\cite[Table 1]{bajcsy2015systematic}. 
Rules for interpreting marginal marks vary across municipalities. The desire to account for voter intent\footnote{\href{https://www.lgbtmap.org/democracy-maps/voter_intent_laws}{Voter Intent Laws Map} shows different guidelines, see for example \href{https://moffatcounty.colorado.gov/sites/moffatcounty/files/voterIntentGuide.pdf}{Colorado's guidelines}.} complicates the question of what images should be in a training set and how they should be labeled.
In all datasets, labels are produced from an optical lens scanner.
Finally, images may be captured as grayscale (8 bpp) or color (RGB, 24 bpp) artifacts. We present four datasets.

\bubblesGray uses 42,679 images ($40\times 50$, 8 bpp)
with blank (35,429 images) and filled (7,250 images) bubbles but no marginal marks. 
\bubblesRGB  a 24 bpp color (RGB) version of \bubblesGray{}. 

\begin{figure}
    \centering
    \includegraphics[width=60mm]{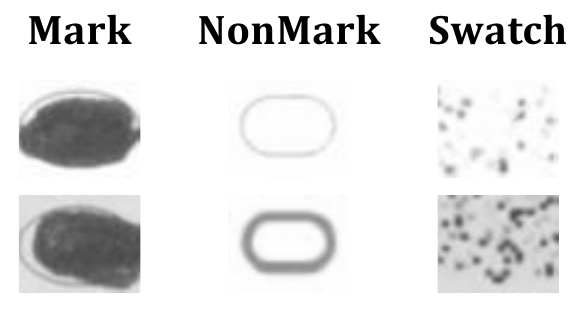}
    \vspace{-2ex}
    \caption[]{Types of examples in our dataset. The swatches are artificial marks designed to be close to the border between a mark and a non-mark for an optical scan. The darker backgrounds are the result of using colored stock paper.}
    \label{fig:bubbleTypes}
    \vspace{-.1in}
\end{figure}

\combinedGray  augments \bubblesGray{} with a collection of marginal marks called ``swatches'' shown in Figure~\ref{fig:bubbleTypes}. Swatches are images that vary the position of signal to create samples close to the boundary of an optical lens scanner. 
The 423,703 randomly generated swatches place equal amounts of random noise throughout each image such that the amount of light is the same. 
This yields 466,382 labeled images.   
\combinedRGB  is a 24bpp color (RGB) version of \combinedGray{}. 

\myparab{Related Voting Datasets}
Two other datasets have been used in voting classification research, these are the Humboldt and Pueblo County Datasets.  The Humboldt county dataset emerged from \href{https://electionstransparencyproject.com/}{Humboldt County Election Transparency Project} but it is {\it not} labeled and we did not have access to a ballot geometry file or a scanner capable of providing labels.  \href{https://county.pueblo.org/clerk-and-recorder-department/ballot-images}{Pueblo County} allows one to access individual ballot images along with the tabulator interpretation.  However, we could not access the entire dataset programmatically.

\subsection{Machine Learning Classifiers}
\label{ssec:classifiers}
This section briefly describes each machine learning model and justifies its inclusion. Details regarding the training and hyper-parameters are given in~\href{https://anonymous.4open.science/r/Busting-the-Ballot}{our anonymous repository}.

\myparab{Support Vector Machine (SVM)}  SVMs support linear classification~\cite{SVM, SVM2}. An SVM maximizes the distance between its decision boundary and inputs of each class label in the training set. We used standard linear kernels to represent a simple classifier. Our SVM has 2,001 trainable parameters for \bubblesGray{} and \combinedGray{} and 6,001 trainable parameters for the RGB datasets.

Non-linear kernels such as RBF (Radial Basis Function) could be used to characterize the boundary of non-linearly separable regions. A non-linear kernel function maps the original data to a higher dimensional space where linear separation may occur~\cite{KernelTrickBook}. Using kernel functions with SVMs is often done on challenging datasets~\cite{elisseeff2001kernel, KaleelXRay}. Exploring the effectiveness of a non-linear kernel such as RBF is future work.

\textit{Why we selected it}: The \emph{linear} SVM represents one of the simplest machine learning models that achieves high accuracy on both the \bubblesGray{} and \bubblesRGB{} datasets. Evaluating the SVM allows us to better understand how robust low-complexity models are to adversarial attacks with voting datasets.

\myparab{SimpleCNN}  Convolutional models are commonly used for image recognition and classification~\cite{lecun1989backpropagation}. SimpleCNN is a shallow convolutional neural network that consists of three identical convolutional layers for a total of 28,818 trainable parameters (grayscale) and 29,394 trainable parameters (RGB).

\textit{Why we selected it:}  SimpleCNN  bridges the gap in complexity between the \emph{linear} SVM and \emph{deep} convolutional neural networks. Its simple architecture provides a lower-bound accuracy for convolution-based models. 

\textbf{Very Deep Convolutional Network (VGG-16). } The VGGNet is a classic convolutional model made to improve AlexNet~\cite{krizhevsky2012imagenet}. The VGGNet architecture restricts the filter size in each convolutional layer to $3 \times 3$. When introduced, VGGNet achieved the largest layer depth when compared to other convolutional models of its time~\cite{simonyan2014vgg}. Our VGG-16 grayscale model has 14,723,010 trainable parameters.

\textit{Why we selected it: } The VGG is the first ``deep" convolution-based neural network. This made the VGG-16 one of the most common benchmarks in traditional image classification~\cite{RESNET, szegedy2015going, szegedy2016inception}.

\myparab{Residual Networks (ResNets)}  Vanilla deep convolutional neural networks are susceptible to  accuracy degradation~\cite{he2015convolutional,srivastava2015highway}.
Residual Networks (ResNets)~\cite{RESNET} offer a solution to this issue. ResNets rely on skip connections between layers. We test a ResNet-20 with 568,033 trainable parameters (grayscale) and 568,321 trainable parameters (RGB). 

\textit{Why we selected it:} ResNets are one of the most widely used types of convolutional neural network. They have been employed in both traditional image classification~\cite{BigTransfer} and in adversarial machine learning extensively~\cite{FAT, BARZ}. 

\myparab{CaiT} Vision transformers are an emerging alternative to convolutional neural networks. Vision transformers benefit from pre-training and their performance excel on image datasets~\cite{VIT}. However, many deep vision transformers suffer from gradient instability and poor feature learning. The Class-Attention in Image Transformers (CaiT) is designed to address these issues in vision transformers~\cite{VIT}. First, a learnable scale parameter is added to regularize residual connections between transformer blocks. Second, CaiT introduces the Class-Attention layer that extracts discriminative features from a class embedding and processed patch embeddings. CaiT has 56,730,626 trainable parameters for our RGB models.

\textit{Why we selected it:} CaiT is one of the state-of-the-art transformer models that has shown excellent performance on image classification tasks. Therefore, it is a natural choice to use as one transformer based alternative to convolutional neural networks.

\myparab{Twins}   The Twins family~\cite{twins} refines the base vision transformer architecture. We test the \texttt{Twins-SVT-B} architecture. It utilizes locally-grouped self-attention along with globally sub-sampled attention to improve the model's performance while solely relying on matrix operation to produce predictions. The Twins model trained on grayscale data has 56,067,880 trainable parameters, whereas for RGB it has 56,070,952 parameters. 

\textit{Why we selected it:} Twins is another representative transformer architecture (like CaiT) with excellent performance on vision tasks.

To summarize, we chose six models across a variety of architectures, linear, convolutional, and attention-based, and sizes, from $2$K parameters to nearly $57$M.

\subsection{ML Classifier Performance on Voting Datasets}

We focus on the performance of grayscale models. We do not observe meaningful variation of trends or results when training on color models.

All six classifiers were trained on the two gray datasets. Table~\ref{table:GrayscaleCombinedAcc} reports the clean training and validation accuracy for the grayscale models. 
Two trends are readily apparent. All models, except the SVM trained on \combined{},  achieve a $99\%$ or greater test accuracy on the easy \bubblesGray{} validation sets, regardless of whether they were trained on \bubblesGray{} or \combinedGray{}.  
When testing on easy marks, all classifiers are  effective, irrespective of their training sets.

Second, high testing accuracy on \combinedGray{} (Combined columns) is not achieved by only training on \bubblesGray{} ({\texttt{-B} rows). The testing accuracies are below $80\%$.  Training on \combinedGray{} (the \texttt{-C} rows) yields better testing accuracies that range from just 59.6\% in grayscale for the SVM  to as high as 93.6\% for Twins. The SVM model is an exception, performance on \combined degrades when trained on combined examples.  We hypothesize the SVM is overfitting its linear boundary to the swatch examples which we believe are close to the true ``boundary'' of the optical scanner.

As mentioned above, an accurate decision boundary for marks other than fully filled bubbles is crucial in real elections. The phenomenon of needing a deep model for high accuracy on complex datasets is consistent with Barretto et al.~\cite{CNNBallot}.

\begin{table}[t]
\small
\begin{center}
\begin{tabular}{l|rr|rr|}
\hline
\multirow{2}{*}{Model (Trained On)} & \multicolumn{2}{c|}{Training Accuracy} & \multicolumn{2}{c|}{Validation Accuracy} \\ \cline{2-5}
                                    & Bubbles & Combined & Bubbles & Combined \\ \hline
SVM-B        & 1.0000 & .5411 & 1.0000 & .7888 \\
SimpleCNN-B  & .9999 & .5776 & .9999 & .7779 \\
VGG-16-B      & 1.0000 & .5351 & 1.0000 & .7859 \\
ResNet-20-B  & .9996 & .5346 & .9998 & .7856 \\
Twins-B      & 1.0000 & .5339 & 1.0000 & .7852 \\ 
CaiT-B      & 1.0000 & .5363 & 1.0000 & .7862 \\ \hline
SVM-C        & .9473 & .7265 & .9171 & .5953 \\
SimpleCNN-C  & 1.0000 & .9077 & 1.0000 & .9170 \\
VGG-16-C      & 1.0000 & .9224 & .9999 & .9300  \\ 
ResNet-20-C  & .9997 & .9896 & .9999 & .9087 \\
Twins-C      & 1.0000 & .9194 & 1.0000 & .9357 \\ 
CaiT-C      & .9995 & .9069&  .9995& .9198 \\ \hline
\end{tabular}
\end{center}
\caption{Clean training and validation accuracies on bubble and combined datasets for grayscale models.}
\label{table:GrayscaleCombinedAcc}
\end{table}

\section{Adversarial Threat Model and Adversarial Example Generation}
\label{sec:adv attacks}
We assume an attacker that compromises ballot printing. 
The attacker delivers the adversarial examples  at the printing stage when blank ballots are either produced, stored in a warehouse, or shipped. The attacker creates printed ballots to be filled out by voters and cast in a tabulator. The attacker has full control of the ballot image and all bubbles must visually {\it appear} to be empty at the onset. As this ``empty'' ballot will be inspected and filled by a voter, the goal is to change the classifier output from non-mark to mark while the added signal remains imperceptible.  

As we discuss in Section~\ref{ssec:attack example}, one does not need to impact tabulator accuracy much to have an impact, Table~\ref{tab:close races} shows the number of close state legislative races in battleground states in the 2020 United States Presidential Election with $15\%$ of races having a margin of less than $5\%$. Our attacks are most harmful when a large number of voters do not specify a preference for the targeted contest. Thus, our attacks are unlikely to impact a presidential race where almost all ballots specify a preference.

\subsection{Attack Nomenclature}
An attacker that compromises ballot printing can only create examples where a mark appears to be empty but will be classified as a mark.  We call this $\advmark$. For completeness, in the virtual domain we also consider an attacker that perturbs a marked bubble so that it is classified as a nonmark. This is called \advnmark as it could lead to a preference being removed and no preference being counted, known as an undervote.

\myparab{Virtual} First in Section~\ref{sec:digital domain}, we consider the idealized (for the attacker) \textit{virtual} context where the attacker modifies an image without any intervening printing or scanning. Namely, the adversarial signal cannot be altered (affected) by the steps taken with real physical ballots. We test both $\advmark$ and $\advnmark$ in this domain.

\myparab{Physical} Second in Section~\ref{sec:physical domain}, the paper considers the more realistic \textit{physical} context where adversarial examples are organized onto sheets of paper which are printed and then scanned using COTS hardware, namely an HP LaserJet-3010 series and a Fujitsu-7600 scanner. 
The resulting images are registered, color-corrected, and segmented (see an overview in Section~\ref{sec:voting overview}).  
Laser printing is a noisy process.  Laserjet printers use dithering to cope with fewer than $256$ greyscale levels and simulate gray intensities.
Commercial offset printing yields less noise.  Yet, tabulators must handle images from commodity printers (such as on-demand ballot printers) as municipalities print ballots when they run out.  In this domain we only test $\advmark$ examples. In the stringent and realistic physical settings where the adversarial signal is printed on paper it is possible to cause misclassification of non-marks to marks at a high enough rate to impact close elections. This work offers evidence that tabulators using  machine learning algorithms are susceptible to adversarial attacks that cause empty bubbles to be interpreted as marks.

\subsection{Analyzing Attacks on Voting Systems}
\label{ssec:attack example}
Attacking voting systems is distinct from conventional adversarial machine learning in two facets. First, a high attack success rate is not required to impact an election. Second, when a misclassification does occur, several different actions can be taken by the voting system. We detail  differences in this subsection.    
When the classifier perceives a \advmark example one of three things occurs:
\begin{enumerate}
\itemsep0em
    \item If the voter did not fill any bubbles in that race (e.g. a local race in a presidential election), the attacker has created a vote for a candidate in that race.  The \href{https://2022voterdata.lci.fsu.edu/}{Leon County Post Election Audit of 2022} found that 19\% of voters leave at least one race blank. In the 2020 Presidential Election in Nevada (a battleground state), 12\% of voters did not vote for their state legislative race. 
    \item If the voter marks the same preference as the adversarial bubble, the choice is aligned with the attacker and there is no impact.
    \item If another bubble is filled in by the voter in the same race, the tabulator should report an \emph{overvote}. The ballot is returned to the voter. The voter then decides whether to submit their ballot anyways or ask for a new ballot to be completed.  In an attack by  Bernherd et al.~\cite{bernhard2021risk} voters do ``not know what to do if they noticed a problem with their paper ballot during a real election.'' In our experience, voters often resubmit their ballots.
\end{enumerate}

\textbf{What attack success rate can flip an election?} In conventional adversarial machine learning, a high robust accuracy (low attack success rate) generally indicates an acceptable defense. For example, one of the most recent state-of-the-art defenses proposed in~\cite{wang2023better} achieved a robust accuracy of $70.69\%$ against white-box adversarial machine learning attacks for the CIFAR-10 dataset. This robustness would correspond to a $29.31\%$ attack success rate. Elections are routinely decided by small margins (Table~\ref{tab:close races}).  An attacker can reuse examples globally, though their reused examples would be subject to scanning noise discussed in Section~\ref{sec:voting overview}.

\begin{table}[t]
\centering
    \begin{tabular}{l | r | r r r| r}
                 2020&  Races&  $5\%$&  $2\%$& $1\%$ & Blank $\%$\\\hline
                 Arizona&  90&  16&  6& 4  & -\\
                 Nevada&  52&  12&  3& 2 & 11.7\%\\
                 Georgia&  236&  14&  5& 2 & 8.9\% \\
                 North Carolina&  170&  15&  4& 0 & 4.7\%\\
                 Pennsylvania &  228&  12&  3& 1 & 7.0\%\\
            \end{tabular}
    \caption{Number of tight state legislative races in United States Battleground States in 2020 Presidential Election. We omit Arizona as each district elects two legislators and many of these districts have at most $2$ candidates.}
    \label{tab:close races} 
\end{table}

We now illustrate how a small attack success rate can impact a close election.
We consider a race with a $2\%$ margin where $12\%$ of ballots are left blank (the rate for Nevada state legislative races in 2020).  We assume a two candidate race. Without an attack, $\mathtt{Win}$ will receive $.415$ fraction of the vote and $\mathtt{Lose}$ will receive $.395$ fraction of the vote.  The adversary's goal is for $\mathtt{Lose}$ to win over $\mathtt{Win}$ by a margin of $.5\%$ (often results under this margin trigger a hand recount). There are three relevant parameters:
\begin{enumerate}
    \itemsep0em
    \item What fraction of ballots carry an adversarial example for the $\mathtt{Lose}$ candidate? We call this parameter $\mathtt{deploy}$.
    \item What fraction of blank ballots with an adversarial example are misclassified? We denote this as $\mathtt{success}$.
    \item If the ballot has an adversarial example that is misclassified as a vote for $\mathtt{Lose}$ and the voter filled in a vote for $\mathtt{Win}$ their ballot will be marked as an overvote. When this occurs what fraction of the time does the voter ask for a new ballot? We call this probability $1-\mathtt{recast}$ and assume in this case  the ballot is counted for $\mathtt{Win}$.  With probability $\mathtt{recast}$, the voter asks the tabulator to accept the overvoted ballot and the ballot is not counted for either candidate.
\end{enumerate}

\noindent
Consider if the parameters are $\mathtt{deploy}=1, \mathtt{success}=.1, \mathtt{recast}=.3$, the votes for each candidate becomes 
\begin{align*}   \mathtt{Lose}&:=.395+.12\cdot \mathtt{deploy}\cdot\mathtt{success} = .407\\
\mathtt{Win}&:=.415(1- \mathtt{deploy}\cdot \mathtt{success}\cdot \mathtt{recast})  = .402
\end{align*}
Some blank ballots are converted to votes for $\mathtt{Lose}$ and some $\mathtt{Win}$ ballots are converted into overvotes which are not counted for either candidate. Looking ahead to Section~\ref{sec:physical domain}, we achieve $\mathtt{success}~\approx~.99$ on the most vulnerable model (a support vector machine); tested model's vulnerability against realistic attacks varies widely (for more resilient models, $\mathtt{success}=0$ for imperceptible examples). 

In the case when a voter refills their ballot after their ballot is marked as an overvote, the second ballot obtained by the voter may also contain an adversarial example that is misclassified as a vote. The overall fraction of ballots where this occurs for the parameters discussed is $.13\%$ of ballots.  Continuing to use Nevada as an example, in 2020 the average state legislative race had 34K cast votes, so $.13\%$ of ballots corresponds to $43$ ballots. Widespread occurrences of a voter having to request a new ballot multiple times is likely to arouse suspicion. This creates an incentive for the attack for $\mathtt{deploy}\cdot \mathtt{success}$ to be less than $1$.

In summary, our analysis of attack success rate for the voting domain reveals a very important issue. In traditional adversarial machine learning a defense is successful if robust accuracy is $70\%$. As the example above shows, one can change the tabulated margin of a race by $2.5\%$ even with a robust accuracy of $90\%$. 

\subsection{Generating Adversarial Examples}
\label{ssec:white box}

Adversarial examples can be created from clean images for a given model either through white-box or black-box adversarial attack methods~\cite{tramer2020adaptive,mahmood2021beware}.
In both, an attacker begins with a clean unperturbed image and injects noise. Throughout this work, all clean images  used to create adversarial examples are from the \bubbles datasets, we \emph{never use a swatch image as a starting point}. This work focuses on white-box attacks due to \href{https://www.eac.gov/voting-equipment/voluntary-voting-system-guidelines}{VVSG 2.0 requirement 1.1.6G}, which tabulators describe methods for classifying marks.

The most common method~\cite{tramer2020adaptive} to create adversarial images adds noise based on gradient information from the model. This is referred to as a white-box attack~\cite{carlini2019evaluating}. In this formulation, the gradient of the input with respect to a certain loss function is computed directly using the target model's architecture and trained weight parameters. This attack is an optimization problem:
\begin{equation}
    \max_{x_{adv}} \mathcal{L}(x_{adv}, y; \; \theta) 
    \mbox{ subject to } ||x - x_{adv}||_p \leq \epsilon
\end{equation}
where $\mathcal{L}$ is a loss function, $x$ is a clean (non-perturbed) image with true class label $y$, $\theta$ represents the parameters of the model being attacked, $\epsilon$ is a bound on the magnitude of the perturbation and $||\cdot||_p$ represents the $l_p$ norm. 
The adversarial example is constrained to be at a distance at most $\epsilon$ from the original clean example. 

We use the $l_{p}$ norm with $p=\infty$ in our attacks. This is a widely used norm in adversarial machine learning~\cite{carlini2019evaluating, mahmood2021robustness, rathbun2022game}. We focus on APGD~\cite{APGD} but some of gradient masking results in the next section use 
PGD~\cite{madry2018towards}.
APGD is a SOTA white-box attack~\cite{mahmood2021robustness, wang2023better, rathbun2022game}. 

\section{Gradient Masking} 
\label{sec:gradient masking evidence}

Gradient masking frequently occurs in adversarial machine learning when evaluating the robustness of defenses to white-box attacks~\cite{athalye2018obfuscated, tramer2020adaptive}. Gradient masking is when the gradient of a model is incorrectly estimated in a white-box attack. This phenomena leads to the model having a falsely high robustness. Often, defenses are proposed and tested with attacks like FGSM and PGD, and are later broken by adaptive attacks which overcome gradient masking~\cite{carlini2017adversarial, antonyan2009state, mahmood2021beware}. Gradient masking does not make a model secure.

For voting datasets, zero gradients occur when backpropagating the gradient in the SimpleCNN and ResNet-20 models during the attack. In addition, we observe non-monotonic behavior of APGD with increasing $\epsilon$ on VGG-16. It is important to note this occurs after the models have been trained. The models often exhibit maximal predictive confidence of either $[1,0]$ or $[0,1]$. In this section, we explore the extent of the issue and show that it is rooted in the numerical instability of floating point and datatypes used by {\tt PyTorch} with NVIDIA GPUs. Furthermore, in Section~\ref{sec:gradient masking}, we show how a modified DLR loss can overcome this issue~\cite{APGD}.

\subsection{The Repeated Zero Gradient Condition}
We demonstrate the occurrence of zero gradients in multiple different models trained on the voting datasets when conducting standard white-box adversarial attacks.

\begin{table*}[t] 
\centering
\begin{tabular}{  l | l | l | r | r | r | r|}
&&& \multicolumn{2}{c|}{Bubbles Validation Set} & \multicolumn{2}{c|}{Swatches Validation Set}\\
Training &&&1st Step & Average & 1st Step & Average\\
Dataset & Model & Class & $0$ Grad & Conf. & $0$ Grad & Conf.\\ 
\hline
\multirow{6}{*}{\bubblesGray} & \multirow{2}{*}{SVM} & Mark & 0 & [0.9939, 0.0061] & 0 & [0.2871, 0.1408] \\ \cline{4-7}
 & & Non-Mark & 0 & [0.1274, 0.8737] & 0 & [0.1512, 0.8488]\\ \cline{3-7}\cline{3-7}
 & \multirow{2}{*}{SimpleCNN} & Mark & 500 & [1.0, 0.0] & 28 & [0.8166, 0.0152] \\ \cline{4-7}
 & & N-Mark & 0 & [1.3808e-21, 1.0] & 0 & [7.009e-21, 1.0] \\ \cline{3-7}\cline{3-7}
 & \multirow{2}{*}{ResNet-20} & Mark & 500 & [1.0, 0.0] & 496 & [0.9819, 2.2076e-6]\\ \cline{4-7}
 & & N-Mark & 500 & [0.0, 1.0] & 500 & [0.0, 1.0]\\ 
\cline{2-7}
 \multirow{6}{*}{\combinedGray} & \multirow{2}{*}{SVM} & Mark & 0 & [1.0, 1.0989e-09] & 0 & [0.2985, 0.1097] \\ \cline{4-7}
 & & N-Mark & 0 & [0.0977, 0.1466] & 0 & [0.0658, 0.1112]\\ \cline{3-7}\cline{3-7}
 & \multirow{2}{*}{SimpleCNN} & Mark & 498 & [1.0, 2.4490e-29]  & 1 & [0.4838, 0.0309]\\ \cline{4-7}
 & & N-Mark & 0 & [0.0, 9.9425e-06] & 0 & [0.0258, 0.7107]\\ \cline{3-7}\cline{3-7}
 & \multirow{2}{*}{ResNet-20} & Mark & 23 & [0.2847, 0.0036]  & 304 & [0.5023, 0.0007] \\ \cline{4-7}
 & & N-Mark & 156 & [0.0020, 0.4903] & 60 & [0.0021, 0.2586]\\ 
\hline
\end{tabular}
\caption{Zero Gradient Condition recorded over 20 steps of PGD. We record the number out of $500$  of examples where zero gradient occurs on the first step, the average number of steps the zero gradient condition occurs in, and the confidence output out of $500$ examples for each class.}
\label{table:ZeroGrad}
\end{table*}

\myparab{Experimental Setup} We attack three models (SVM, SimpleCNN, ResNet-20) trained on the grayscale datasets (\combinedGray and \bubblesGray) using \pgd. We set \pgd to run for 20 steps using $\epsilon = 0.031$ with step size $0.00155$. We randomly select 500 marks and 500 non-marks from the \bubbles validation set (no swatches) and 500 marks and 500 non-marks from the \emph{Swatches only} (no bubbles) validation set that were correctly classified. At each of the $k$ steps of \pgd we check the maximum element of the absolute value of the gradient matrix. If  $\max_{i}\left\{\left|{\partial L/\partial x^{(k)}_{i}}\right|\right\}=0.0$ then this step of \pgd exhibits a zero gradient. 

\myparab{Analysis of Zero Gradient} The number of recorded instances of zero gradient across 20 steps are reported in Table~\ref{table:ZeroGrad}. All 500 marks encounter a zero gradient for the first step on the SimpleCNN. All 500 marks and 500 non-marks encounter a zero gradient for the first step on ResNet-20. No example for the SVM encounters a zero gradient in the bubbles validation set. Note that while fewer swatch examples express a zero-gradient, only bubbles are considered a valid starting point for our attacks.

\myparab{Analysis of Confidence} Our models return a tuple for their confidence in each class, vote then non-vote respectively. We provide the average confidence tuple (rounded to four decimal places) over each step for each class. Most notably, for classes and models that encounter a zero gradient for all 500 examples, the confidence is either $[1.0, 0.0]$ for votes or $[0.0, 1.0]$ for non-votes. Since each model uses a {\tt softmax} activation layer to normalize their outputs, a $1.0$ is the maximum possible confidence for a class.

\subsection{Numerical Instability}
We devise an experiment that attributes zero-gradient to numerical instability arising from 32-bit floating point arithmetic as well as tensor-float arithmetic in use within the {\tt PyTorch} implementation. We first introduce some notation on our target models. 

\myparab{Confidence and Gradient} A machine learning classifier receives an image as input $x$, forward propagates it through multiple layers $L_i (x) \rightarrow L_{i+1}(x)$, then outputs a confidence vector at the final layer. This output is the prediction this image belongs to each class $\tilde{y}$. The loss function $\mathcal{L}$ derives how far the prediction $\tilde{y}$ is from ground truth label $y$. 

White-box adversarial machine learning attacks use gradient descent on the loss $\mathcal{L}$ \emph{with respect to the source image\footnote{This differs from gradient descent during training that computes the derivative with respect to network weights.}} $x$, i.e.,  $\frac{\partial \mathcal{L}}{\partial x}$, the gradient at the network input.  
A zero gradient $\frac{\partial \mathcal{L}}{\partial h}=0$ appearing at some layer during the backpropagation will spread to shallower layer and induce 0 gradients all the way to the input layer, that is, all the way to $\frac{\partial \mathcal{L}}{\partial x}$.  
We observed zero gradient at the final {\tt softmax} layer. 
We next review the specifics about floating point arithmetic to help understand the root cause.

\myparab{Floating point refresher} IEEE-754 is the standard defining 32-bit floating point numbers. The use of fixed precision (32-bits $b_{31}b_{30}\cdots b_1b_0$), implies only certain numbers are representable. The standard calls for 1 bit for the sign $b_{31}$, 8 bits for the exponent $b_{30}\cdots b_{23}$ (using a bias representation) and 23 bits for the mantissa $b_{22}\cdots b_0$ to encode a \emph{normal} floating point value:
\[
(-1)^{b_{31}} \cdot 2^{\left(\sum_{i=23}^{30} b_i \cdot 2^{i-23}\right) - 127} \cdot 
\left(1 + \sum_{i=1}^{23} b_{23-i} \cdot 2^{-i}\right)
\]
This representation uses an implicit $1$
at the start of the mantissa. The smallest normal floating point is 
\ifnum\conference=0
\[
(-1)^0 \cdot 2^{-126} \cdot (1 + 0) = 2^{-126} \approx 1.1754943508 \cdot 10^{-38}.
\]
\else
$\approx 1.1754943508 \cdot 10^{-38}$.
\fi
The range of representable floats can be \emph{extended with de-normalized representations} where the first mantissa bit is zero. This broadens the range by using 0 bits at the most-significant end of the mantissa to boost the exponent at the expense of the number of digits of accuracy. The smallest de-normalized 32-bit floating point is $1.401298 \cdot 10^{-45}$ which, in binary, is all zeroes except the least significant bit of the mantissa. 
To retain accuracy, computed values should never drop past the smallest \emph{normal} floating point.

\myparab{PyTorch Floating Points} The most popular ML framework {\tt PyTorch} uses the TensorFloat-32 ({\tt TF32} representation for floating point numbers supported by the \href{https://blogs.nvidia.com/blog/tensorfloat-32-precision-format/#:~:text=TF32%20uses%20the%20same%2010,learning%20and%20many%20HPC%20apps.}{NVIDIA hardware} to compute convolutions). 

This shorter representation uses only 10 bits of mantissa, 8 bits of exponents and a sign bit. They are designed \emph{for speed} (about an order of magnitude faster) and are considered \emph{good enough} for the precision expected by machine learning. 
\urldef{\tfurl}\url{https://pytorch.org/docs/stable/notes/cuda.html#tf32-on-ampere}
The flags are, respectively, {\tt torch.backends.cuda.matmul.allow\_tf32} 
to enable them for matrix multiplications and {\tt torch.backends.cudnn.allow\_tf32} to
enable them for convolutions. They were introduced in {\tt PyTorch}
version 1.7.\footnote{PyTorch TF-32 \tfurl.}
By default, CNNs use {\tt TF32} for
key computations during forward and backward passes.\footnote{There is no reason to expect this computation to match a sequential CPU computation as floating points are not commutative. GPUs will group operands differently and results may change.}

\myparab{Manual Backpropagation} Models here contain a final linear layer that feeds into a {\tt softmax} activation function. A linear layer accepts a feature vector $h$, performs matrix multiplication with weight matrix $W$, then adds a bias term $b$. The $i^{th}$ column in the output $z$ is the confidence the input image $x$ belongs to $(i-1)^{th}$ class. The {\tt softmax} layer exponentiates each column in $z$ then normalizes them over their sum, returning the confidence vector $\tilde{y}$.
\[z = h W^T + b \quad \rightarrow \quad \tilde{y} = \dfrac{e^z}{\sum_{i=1}^C e^{z_i}}\]
This allows to evaluate the CE Loss. Note that $y$ is the one-hot encoding of  the image's class.
\[\mathcal{L} = -\sum_{i=1}^2 y_i \cdot \log(\tilde{y}_i)\]
Consider the gradient of this loss with respect to a feature vector $h$. We can express it w.r.t. the output of the linear layer $z$:
\[\dfrac{\partial \mathcal{L}}{\partial h}  = \dfrac{\partial \mathcal{L}}{\partial z} \cdot \dfrac{\partial z}{\partial h}\]
\noindent Since $z = h W^T + b$, the derivative w.r.t. $h$ is just the weight matrix $\dfrac{\partial z}{\partial h} = W^T$ and the derivative of the loss w.r.t. $z$ is  $\dfrac{\partial \mathcal{L}}{\partial z} = \tilde{y} - y$. The product of these terms delivers the full backpropagation equation. 

\myparab{Backpropagation Experiments} 
Both the accuracy of 32-bit floating point and the reduced accuracy of the TF-32 type contribute to zero gradients, indeed, the calculations of $z$ and $\tilde{y}$ involve convolutions in {\tt PyTorch}. 

\begin{figure}[t]
    \centering
    \includegraphics[width=60mm]{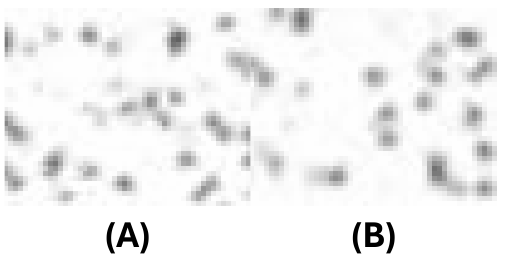}
    \caption{Examples of mark swatches considered for manual backpropagation. The ResNet-20 produces a zero gradient for Swatch (A) and a non-zero gradient for Swatch (B).}
    \label{fig:gradSwatches}
\end{figure} 
\myparab{32-bit Floating point accuracy} Consider two swatches $A$  and $B$ shown in Figure~\ref{fig:gradSwatches} that are members of the same class (their $y$ vectors are  $[1,0]$). We chose mark swatches because of their visual similarity and they can exhibit
zero gradients (see Table \ref{table:ZeroGrad}).
Empirically, $A$ triggers a zero-gradient while $B$ does not. 
Given the \emph{fixed} weights of the ResNet-20, we can manually compute  $z$  and  $\tilde{y}$ using the penultimate layer weights $W$ and biases $b$. Namely:
\[
z(A) = h(A) \cdot W^T + b \:,\: z(B) = h(B) \cdot W^T + b 
\]
as well as 
\[
\tilde{y}(A) = \frac{e^{z(A)}}{\sum_{i=1}^C e^{z_i(A)}} 
\:,\: \tilde{y}(B) = \dfrac{e^{z(B)}}{\sum_{i=1}^C e^{z_i(B)}} 
\]
Those values are used to compute $\dfrac{\partial \mathcal{ L}}{\partial h} = (\tilde{y} - y) \cdot W$ for both $A$ and $B$. 
To understand the stability issue, consider the following values for $z(A)$ and $z(B)$
\begin{align*}
z(A) &=& [49.218 \:\:\: -48.582] \:,\: 
z(B) &=& [18.516 \:\:\: -18.059]
\end{align*}
The last layer contains 2 neurons, so we get two z-values. Computing $\tilde{y}(A)$ produces
\begin{align*}
e^{z(A)} = [e^{z_1(A)} e^{z_2(A)}] &= [e^{49.218}\hspace{1em} e^{-48.582}] \\&= [2.3720\cdot 10^{21}\hspace{1em}  7.9635\cdot 10^{-22}]
\end{align*}
To get $\tilde{y}(A)$, we compute $e^{z_1(A)}+e^{z_2(A)}$ as 
$2.3720\cdot 10^{21}$, i.e., $e^{z_1(A)}+e^{z_2(A)} = e^{z_1(A)}$. The magnitude difference between $e^{z_1(A)}$ and $e^{z_2(A)}$ is so large  that the second operand is \emph{absorbed} by the first. The second ratio \[\tilde{y}_2(A)=\dfrac{e^{z_2(A)}}{e^{z_1(A)}+e^{z_2(A)}}=0\]  because the division of a very small number by a very large one underflows the float type. Overall, $\tilde{y}(A) = [1 \: 0]$ and the first factor of the gradient is $\tilde{y}(A) - y(A) = [1 \: 0] - [1 \: 0] = [0 \: 0]$. Interestingly,
with $B$, the $z$ values are a bit smaller leading to 
\[
e^{z(B)} = [ 1.1000\cdot 10^{8} \hspace{1em} 1.4357\cdot 10^{-8}]
\]
and $\tilde{y}(B) = [9.999999\cdot 10^{-1} \hspace{1em}  1.305207\cdot 10^{-16}]$ which does not trigger the zero gradient. The gradient expressions above were manually derived and evaluated with Octave~\cite{octave} to independently confirm the observations. 
\emph{With an FP32 representation, the backpropagation through the last layer can yields a zero gradient.  Once this occurs, the preceding gradients will be $0$ as well.}

\myparab{TensorFloat 32-bit Floating point accuracy} 
Recall that {\tt PyTorch} uses convolutions to compute $z$ and $\tilde{y}$. Given the defaults used by {\tt PyTorch}, these convolutions do rely on the numerically weaker TF32 type (10-bit mantissa). The accuracy of the $z$ values and the $\tilde{y}$ values are further reduced. This can increase the occurrence of zero gradients for the same reasons.

Since zero gradients are a direct consequence of the data types used within {\tt PyTorch} for our datasets, we believe this phenomenon is likely to occur on bubble classifiers produced by industry and researchers. We note that while the problem could be made worse by the 
specialized datatypes with less precision on NVIDIA GPUs, the problem \emph{still} occurs with classic 32-bit floating point. Turning off these datatypes has the undesirable effect of causing a $10\times$ slowdown during training. We have not tested behavior on models using $64$-bit wide floating points.  

One potential solution for the attacks is to employ a randomized start (commonly done in \pgd and APGD). However, given that zero gradients are so frequent, they can \emph{still} occur after the first step of the attack.  In addition, randomized start does not provide a deterministic solution to the problem.

\section{Overcoming the  Zero Gradient Condition}
\label{sec:gradient masking}
The zero-gradient condition has previously been encountered~\cite{athalye2018obfuscated} when assessing adversarial machine learning defenses. In our work gradient masking (zero gradients) occurs in the models trained on the ballot datasets, without any defenses implemented. To the best of our knowledge, we are the first to observe this phenomenon in classifiers without defensive mechanisms. We show how to overcome this issue using a modified version of the difference of logits ratio (DLR) function proposed in~\cite{APGD}. 

As an alternative to using cross-entropy (CE) loss, in~\cite{carlini2017towards} the Carlini and Wagner targeted attack was proposed in which the following loss function was minimized:
\begin{equation}
\label{eq:carlini}
F(x) = \text{max}(z(x)_{t}-\text{max}\{(z(x)_{j}:j \neq t\}, -\kappa)
\end{equation} where ${z(\cdot)_{j}}$ is the ${j^{th}}$ logit output from the model, ${z(\cdot)_{t} }$ represents the logit of the target class $t$ and $\kappa$ represents confidence with which the adversarial example should be misclassified. Further work in~\cite{APGD} proposed the use of DLR loss function:
\begin{equation}
\label{eq:DLR}
    \text{DLR}(x, y) = -\frac{z_y - \max_{j\neq y} z_{j}}{z_{\pi_1} - z_{\pi_3}}
\end{equation}
where $z_{y}$ is the logit output corresponding to the correct class label and $\pi$ is a permutation that orders the elements of the logit output $z$ in decreasing order. It is important to note that the DLR loss function is for multi-class classification where the number of classes $C$ is greater than 3, since $z_{\pi_3}$ is undefined for $C<3$. However, the ballot datasets are binary classifications tasks ($C=2$). Hence we modify the DLR loss function in Equation~\ref{eq:DLR} by only using the numerator. Effectively this reduces DLR loss function to the untargeted version of the Carlini and Wagner loss function introduced in Equation~\ref{eq:carlini}, without the outer maximization and confidence $\kappa$. It is important to note that the denominator $z_{\pi_1} - z_{\pi_3}$ in Equation~\ref{eq:DLR} was included for scale invariance to prevent gradient masking~\cite{APGD}. In our experiments, we observe that the binarized DLR loss operates as expected despite removing the denominator.

\section{Adversarial Attacks in the Virtual Context}
\label{sec:digital domain}
The virtual context is a best-case scenario for an adversary. When crafting perturbations, we ignore artifacts (e.g., noise) introduced by the physical world. Performance in the more challenging \emph{physical context} appears in Section~\ref{sec:physical domain}.

Figure~\ref{fig:attackType} shows how bubble images yield two different types of attacks. Recall that $\advmark$ are  adversarial marks that appear empty yet are classified as marks. Likewise,  $\advnmark$ are adversarial mark that appear to be marks but are classified as blanks. Recall $\advnmark$ cannot be conducted by an attacker compromising a print vendor but are presented for completeness.
The first set of experiments is designed to answer fundamental security questions:
\begin{enumerate}
\itemsep0em
    \item Which attacks are most effective? 
    \item Does the $DLR$ loss overcome the zero-gradient condition?
    \item Does the training dataset impact model resilience? 
    \item Does model complexity impact attack success rate?
    \item Do imperceptible $\epsilon$ alterations flip the classifier output?
\end{enumerate}

\begin{figure}[t]
    \centering
    \includegraphics[width=82mm]{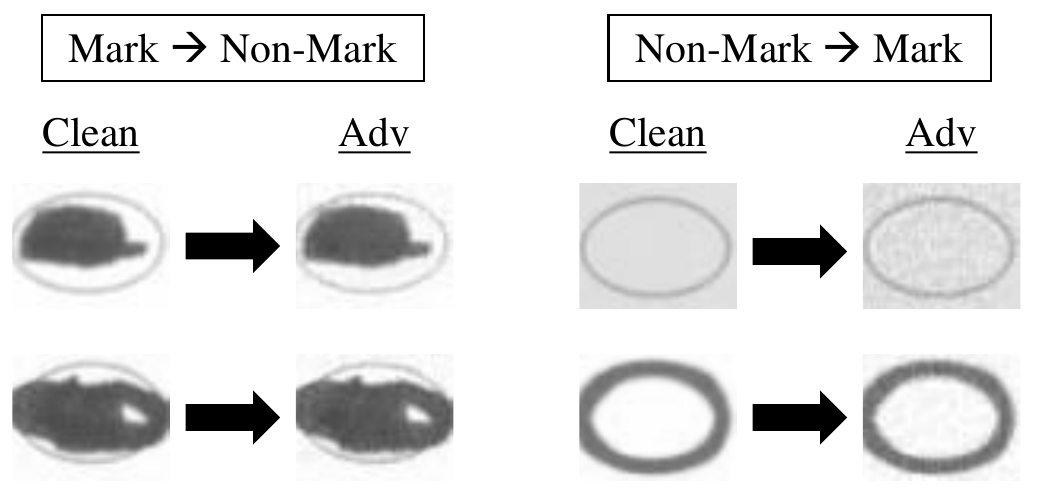}
    \vspace{-2ex}
    \caption[]{Examples of mark to non-mark and non-mark to mark adversarial attacks on the SimpleCNN. We abbreviate mark $\rightarrow$ non-mark as \advnmark and non-mark$\rightarrow$ mark as $\advmark$. The examples above are created with APGD $\epsilon=8/255$.}
    \label{fig:attackType}
    \vspace{-.1in}
\end{figure} 

\subsection{White-Box Attack Performance}
We first investigate the choice of attack. We use model robustness at each perturbation magnitude ($\epsilon$ value) to determine the best performing attack.

\myparab{Experimental Setup} We attack six models (SVM, SimpleCNN, VGG-16, ResNet-20, CaiT, and Twins) trained on the grayscale datasets (\combinedGray and \bubblesGray) using \apgd. We consider two versions of \apgd 1) with CE loss, 2) with DLR loss.  

We randomly select 500 marks and 500 non-marks from the \bubbles validation set (no swatches) that were correctly classified by the target model. The choice of clean initial samples follows Mahmood, Mahmood, and van Dijk's methodology~\cite{mahmood2021robustness}. For all attacks, $\epsilon$ varies from $4/255$ to $255/255$.  We report the resulting robust accuracy in Table~\ref{table:varyEps} for the \combinedGray and \bubblesGray datasets.

\begin{table*}[t]
    \centering
    \begin{tabular}{l | l | l | r || r|r|r|r|r|r|}
        Model & Dataset & Loss & Clean Acc. & 4/255 & 8/255 & 16/255 & 32/255 & 64/255 & 255/255 \\\hline

        \multirow{4}{*}{SVM}
            & \combinedGray & CE & \multirow{2}{*}{.9171} & .500 & .500 & .500 & .499 & 0 & 0 \\
            & \combinedGray & DLR &  & .500 & .500 & .500 & .499 & 0 & 0 \\\cline{4-10}
            & \bubblesGray & CE & \multirow{2}{*}{1.0000} & 1.000 & 1.000 & 1.000 & 1.000 & .138 & 0 \\
            & \bubblesGray & DLR &  & 1.000 & 1.000 & 1.000 & 1.000 & .138 & 0 \\\hline

        \multirow{4}{*}{SimpleCNN}
            & \combinedGray & CE & \multirow{2}{*}{1.0000} & 1.000 & .825 & .500 & .498 & .498 & .500 \\
            & \combinedGray & DLR &  & 1.000 & .824 & .500 & 0 & 0 & 0 \\\cline{4-10}
            & \bubblesGray & CE & \multirow{2}{*}{.9999} & 1.000 & 1.000 & 1.000 & 1.000 & 1.000 & .500 \\
            & \bubblesGray & DLR & & 1.000 & 1.000 & 1.000 & 1.000 & .584 & 0 \\\hline

        \multirow{4}{*}{VGG-16}
            & \combinedGray & CE & \multirow{2}{*}{.9999} & 1.000 & 1.000 & .492 & .020 & 0 & .453 \\
            & \combinedGray & DLR &  & 1.000 & 1.000 & .495 & .023 & 0 & 0 \\\cline{4-10}
            & \bubblesGray & CE & \multirow{2}{*}{1.0000} & .984 & .134 & 0 & 0 & 0 & 0 \\
            & \bubblesGray & DLR &  & .984 & .137 & 0 & 0 & 0 & 0 \\\hline

        \multirow{4}{*}{ResNet-20}
            & \combinedGray & CE & \multirow{2}{*}{.9999} & .168 & .304 & .647 & .647 & .645 & .644 \\
            & \combinedGray & DLR &  & 0 & 0 & 0 & 0 & 0 & 0 \\\cline{4-10}
            & \bubblesGray & CE & \multirow{2}{*}{.9998} & 1.000 & 1.000 & 1.000 & 1.000 & 1.000 & 1.000 \\
            & \bubblesGray & DLR &  & 1.000 & 1.000 & .995 & .921 & .500 & 0 \\\hline

        \multirow{4}{*}{CaiT}
            & \combinedGray & CE & \multirow{2}{*}{.9995} & 1.000 & 1.000 & .500 & .499 & .090 & 0 \\
            & \combinedGray & DLR &  & 1.000 & 1.000 & .500 & .499 & .092 & 0 \\\cline{4-10}
            & \bubblesGray & CE & \multirow{2}{*}{1.0000} & 1.000 & 1.000 & 1.000 & .991 & .501 & 0 \\
            & \bubblesGray & DLR &  & 1.000 & 1.000 & 1.000 & .991 & .501 & 0 \\\hline

        \multirow{4}{*}{Twins}
            & \combinedGray & CE & \multirow{2}{*}{1.0000} & .243 & 0 & 0 & 0 & 0 & 0 \\
            & \combinedGray & DLR &  & .011 & 0 & 0 & 0 & 0 & 0 \\\cline{4-10}
            & \bubblesGray & CE & \multirow{2}{*}{1.0000} & .979 & .829 & .303 & 0 & 0 & 0 \\
            & \bubblesGray & DLR &  & .979 & .828 & .282 & 0 & 0 & 0 \\\hline
    \end{tabular}
    \caption{Robust Accuracy Results under \apgd\ attack across models using CE and DLR losses on both \combinedGray\ and \bubblesGray\ datasets.}
    \label{table:varyEps}
\end{table*}

\myparab{Analysis of Attack Performance} Model robustness across all grayscale datasets are reported in Table~\ref{table:varyEps}. We additionally tested prior attacks of FGSM~\cite{FGSMpaper}, MIM~\cite{MIM}, 
and PGD~\cite{madry2018towards}. In Appendix~\ref{app:other attacks}, we show that all attacks on ResNet-20 using CE exhibit a non-zero robustness with $\epsilon=255/255$ giving indication the zero-gradient condition occurs for all standard attacks (Table~\ref{table:varyEps resnet}).

\myparab{Training Dataset Matters}
Usually, models trained on the \bubblesGray dataset deliver more robust accuracies than their counterparts trained on the \combinedGray dataset. The only exception to this is the VGG-16 model which is much more robust when trained on $\combinedGray$. The \bubblesGray SVM and SimpleCNN are completely robust up to $\epsilon = 32/255$, CaiT is completely robust until $\epsilon=16/255$, and ResNet-20 is completely robust until $\epsilon = 8/255$. 

The more challenging and realistic \combinedGray training dataset conveys a different picture. Indeed, the SVM classifier accuracy drops to $50\%$ with the smallest $\epsilon=4/255$ while ResNet-20 and Twins drop much further, even for small values of $\epsilon$. SimpleCNN and VGG-16 retain some resilience at small values of $\epsilon (\leq 8/255)$.

As expected, for the majority of the models, training on a more complex dataset that forces the classifier to learn marginal marks, thus moving the decision boundary in a way that makes adversarial attacks easier. The only exception to this rule is VGG-16 which we hypothesize is due to the VGG architecture. The literature has shown the VGG family of models to have other intriguing adversarial properties~\cite{su2018robustness}. It is also worth noting that while VGG-16 trained on \combinedGray is more robust than its bubbles counterpart, VGG-16 is not the most robust CNN model.

From a functionality standpoint though, tabulators must handle marginal marks. Recall from the Introduction and Table~\ref{table:GrayscaleCombinedAcc} that training on $\bubbles$ reduces the clean accuracy of models tested on $\combined$ by $12-15\%$ drastically impacting the accuracy of the classifier on marginal marks. As a reminder, SVM actually increases performance on \combinedGray by training on \bubblesGray, but is not accurate enough for practice when trained using either dataset. {\it One cannot sacrifice performance on marginal marks for resilience to adversarial examples.}  

When compared to the first row of Table~\ref{table:varyEps}, all models trained on \combinedRGB, except ResNet-20, achieved similar or worse robustness than their \combinedGray counterparts up to $\epsilon = 32/255$. We attribute this to more image channels creating a higher dimension space that is easier to exploit. As we discuss in Section~\ref{sec:physical domain}, grayscale images are the preferred method in modern tabulation equipment.

\myparab{Analysis of Model Complexity} Madry et al. argued that increasing model complexity increases robustness to single step adversarial machine learning attacks~\cite{madry2018towards}. There does not seem to be a connection between model complexity and robust accuracy in our results.  Tables~\ref{table:varyEps}  shows that Twins and ResNet-20 are the most vulnerable models while SimpleCNN and CaiT are the most resilient.

\subsection{White-Box Perturbation Magnitude}
We now investigate the choice of $\epsilon$. A large $\epsilon$ yields noticeable adversarial perturbations. In conventional adversarial machine learning there is generally a monotonic relationship between robustness and $\epsilon$ i.e., increasing $\epsilon$ decreases robustness. However, for the our datasets, CE \apgd does not exhibit monotonic behavior for the CNN models as shown in Table~\ref{table:varyEps}. As discussed above, this is due to the difficulty in gradient estimation for these datasets. As an extreme example, VGG-16 on $\combinedGray$ with $\epsilon =64/255$ has model robustness of $0$ but $\epsilon =255/255$ increases model robustness to $.453$. However, DLR \apgd does exhibit monotonic behavior for all models.  

\begin{figure}[t]
\begin{center}
\includegraphics[width=0.45\textwidth]{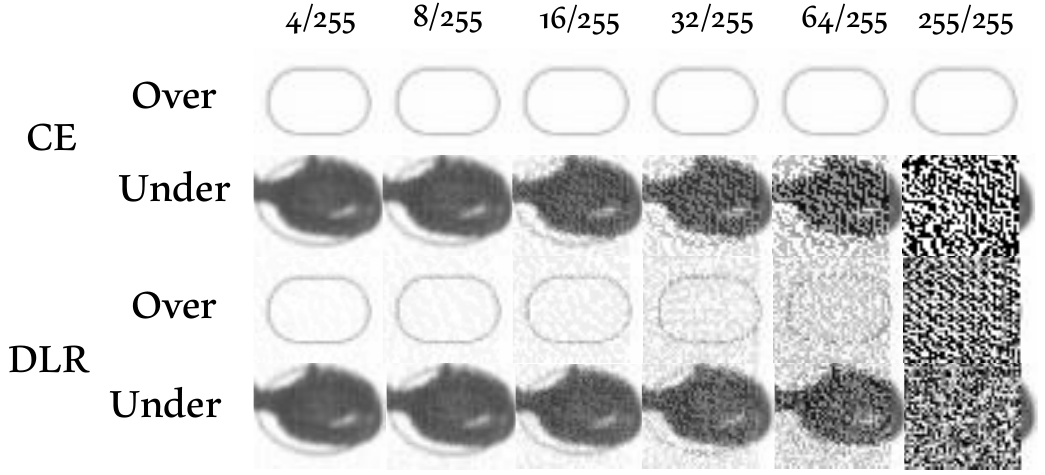} 
\end{center}
\caption{Adversarial examples from varying $\epsilon$ for APGD variants on ResNet-20 trained on \combinedGray dataset. Note that the \advmark example in APGD-CE experiences gradient masking.}
\label{fig:varyEpsFig Loss}
\end{figure}

\myparab{Analysis of Attack Perturbation}
Figure~\ref{fig:varyEpsFig Loss} shows attack images for varying $\epsilon$ when considering CE and DLR.  The model under attack is ResNet-20 trained on $\combinedGray$.  We note the lack of increasing noise in the first row indicating gradient masking.

We now consider Figure~\ref{fig:varyModelFig} which shows increasing $\epsilon$ for each model using the DLR loss. 
For $\epsilon=16/255$ the attack signal starts to be noticeable for all of our attacks. It is also worth noting that in Table~\ref{table:varyEps}, \combinedGray SimpleCNN has robust accuracy of 0.5  at $\epsilon=16/255$. At this point $\advmark$ examples have all flipped class labels (without any changes of $\advnmark$ labels).  
Accuracy restricted to $\advmark$  is shown in Table~\ref{tab:physical success dlr} as that is the focus of our physical world experiments. All six models have robustness of $0$ for \advmark examples at $\epsilon=16/255$. Attacks are easily detectable at the next noise level of $\epsilon= 32/255$. This noise level  is when our attacks start creating \advnmark examples for all models.  

We consider attacks under $8/255$ unnoticeable in the virtual domain and $16/255$ unnoticeable in the physical domain, see Section~\ref{sec:physical domain}.

\begin{figure}[t]
\begin{center}
\includegraphics[width=0.47\textwidth]{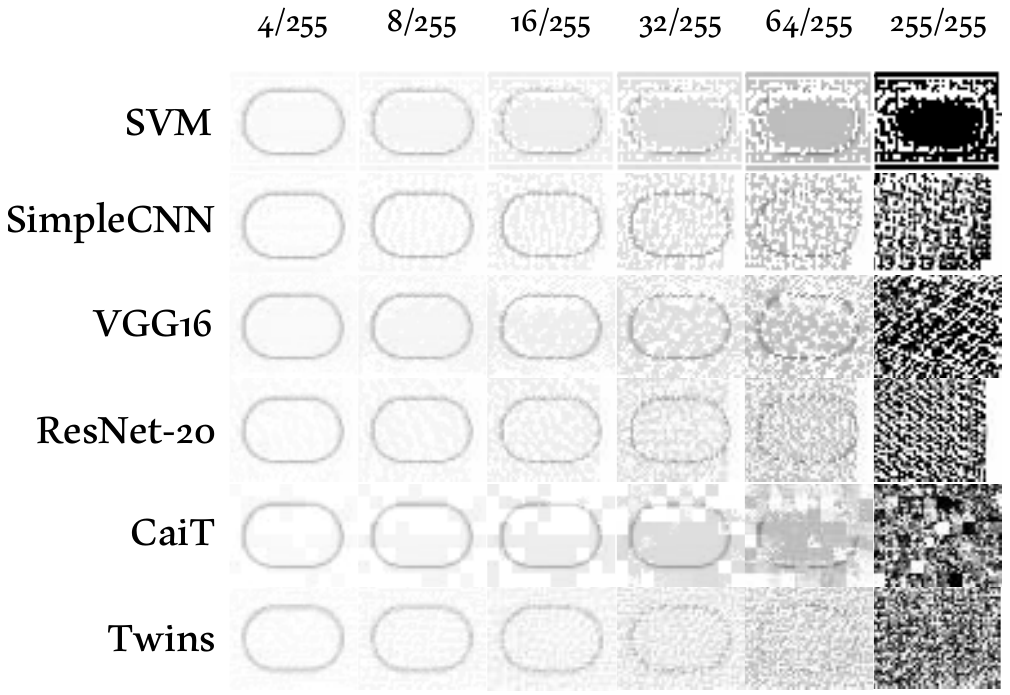} 
\end{center}
\caption{Adversarial examples from varying $\epsilon$ for \apgd with DLR on models trained on \combinedGray dataset.}
\label{fig:varyModelFig}
\end{figure}

\myparab{Visibility of Examples Across Models}
Figure~\ref{fig:varyModelFig} shows the impact of $\epsilon$ when attacking models trained on \combinedGray using \apgd. Only \advmark examples are shown since this represents the more successful attack setting in the physical domain. Figure~\ref{fig:varyModelFig} shows that for all models, $\epsilon \le 8/255$ produces imperceptible perturbations. These perturbations become noticeable when $\epsilon \ge 16/255$. The amount of noise seems to qualitatively be the largest for the SVM and CaiT models. We also note attention artifacts appear clearly visible in attacks on the CaiT  model.  The noise level  for the SimpleCNN, VGG-16, ResNet-20, and Twins models appears similar. 

\section{Attacks in the Physical World}
\label{sec:physical domain}

In this section, we consider the \emph{Print} threat model in the physical world. This attack scenario uses an adversarial printer to generate $\advmark$ examples on printed ballots. We start by reviewing prior work on physical world adversarial machine learning.

\myparab{Prior Work}
Wei et al.~\cite{wei2024physical} surveyed adversarial machine learning in the physical world. Sharif et al.~\cite{Sharif2016} designed targeted adversarial perturbations using a modified softmax loss function. They printed these $224\times 224$ pixel images, of which approximately 6\% area is covered by an adversarial patch in question that takes the shape of glasses. These adversarial glasses were then used to attack (evade) facial recognition systems. This work can be considered human-perceptible perturbations. Recall, that in our attacks, one must produce ballots that are indistinguishable from empty ballots. Kurakin, Goodfellow, and Bengio~\cite{Kurakin2017b} use a modified version of FGSM to design untargeted perturbations at various degrees of $\epsilon$ and print clean and adversarial images, and classify using a phone camera. They use the CE loss function. These adversarial images are human imperceptible.  Like our work, they also consider $l_\infty$ distance. However, their work does not consider printer dithering that we now discuss.

\myparab{Our Pipeline}
We focus on $\advmark$ examples that can be created by an adversarial printer. Attacks follow the following physical pipeline: 

\begin{enumerate}
\itemsep0em
    \item Adversarial example generation (see Section~\ref{sec:digital domain}).
    \item Layout on an empty page.
    \item Printing using a commodity laserjet printer.
    \item Scanning using a commodity scanner.
    \item Image alignment, color correction, segmentation.
    \item Classification using the target model.
\end{enumerate}
Steps 2-5 are not needed in Section~\ref{sec:digital domain}.  This attack is not perfectly realistic in the following sense:
\begin{enumerate}

\item A ballot printer could use higher-end printing techniques (e.g., offset printing, or even photo-realistic printing) instead of laserjet printing.     
\item We use ad hoc registration and segmentation.  Yet, it appears to not introduce measurable/substantial error.
\end{enumerate}

 Vendors have moved away from \emph{colored} ballots in favor of \emph{black and white} ballots where colors are only decorative. 
This section further restricts the investigation to models trained on the \combined dataset to understand the impact of the physical world on models that could realistically serve in place of optical scanners.

\ifnum\conference=0
\subsection{Physical Dataset}
\fi
All the models are trained as described in Section~\ref{sec:dataset_classifiers}. However, physically extracted bubbles are first fed through a denoising autoencoder to remove the noise introduced in the printing process, then fed through a classifier. Without this denoiser, models had clean accuracy of under $90\%$ on bubbles, see Table~\ref{table:CleanAccPrinting} in the Appendix.  The denoiser is described in Appendix~\ref{ssec:denoiser}. In comparison to previous physical attacks, we see substantial noise in classifying clean images when printing and scanning, due  to 1) dithering of the printer and 2) printing a limited set of pixel values. 

\myparab{Layout}
We layout all bubbles ($40 \times 50$ at $200$ dpi) in a matrix on a $8.5 \times 11$ inch sheet spread $0.5$ inches apart. These sheets are  printed and scanned. Registration corrects misalignment from the scanner, and extracts each bubble from their exact locations. 

\myparab{Printing}
We used an HP LaserJet-3010 printer. It is a monochrome laser printer with max print speed of 42 pages per minute (ppm). It prints at 1200 dots per inch (dpi). 

\myparab{Scanner} The scanner is a Fujitsu-7600 with 24bpp and a standard automatic document feeder. We scanned in grayscale at 200 dpi. The software used is \href{http://www.sane-project.org/}{SANE}. The dpi of the printer and scanner are multiples of each other (6x) to avoid fractional scaling.

\begin{table*}[th!] 
    \centering
\begin{tabular}{ l |l | r || r r ||r ||r r r|}
 \multirow{2}{*}{Model} & \multirow{2}{*}{Domain}  &  \multicolumn{7}{c|}{Perturbation Magnitude}\\\cline{3-9}
 && 0/255 & 4/255 & 8/255 & 16/255 & 32/255 & 64/255 & 255/255\\\hline
 \multirow{2}{*}{SVM} &Digital &.874 & 0 & 0 &0 &0 &0 &0\\
  & Phys. & .8480 & .984 & .226 & 0 & 0 &0 &0\\\cline{3-9}
  \multirow{2}{*}{SimpleCNN} &Digital &1.000 & 1.000 & .450 & 0 & 0 &0 &0\\
  & Phys. & 1.0000 & 1.000  & 1.000 & 1.000 & 1.000  &1.000  &0\\\cline{3-9}
  \multirow{2}{*}{VGG-16} & Digital & 1.0000 & 1.000 & 1.000 & 0 & 0 &0 &0 \\
  & Phys. & 1.0000 & 1.000 & 1.000 & 1.000 & 1.000 &.812 & 0\\\cline{3-9}
   \multirow{2}{*}{ResNet-20} & Digital & 1.000 & 0 & 0 & 0 & 0 &0 & 0\\
  & Phys. &0.9940 & 1.000 & .984 & .962 & .996 & .804 & 0\\\cline{3-9}
   \multirow{2}{*}{Twins} &Digital & 1.0000 & .044 & 0 & 0 & 0 & 0 &0 \\
  & Phys. & 1.0000 & .842 & .842 & .844 & .842 & .836 & 0\\ \cline{3-9}
   \multirow{2}{*}{CaiT} & Digital &1.0000 & 1.000 & 1.000 &0 &0 &0 &0\\
  & Phys. & 1.0000 & 1.000 & .974 & .136 & 0 & 0 & 0\\\hline
  
\end{tabular}
\caption{Model robustness against 500 pre/post-print $\advmark$ and 500 pre-print examples generated using \apgd with DLR loss. $\epsilon=8/255$ is deemed imperceptible in the digital domain and $\epsilon=16/255$ in the physical domain. 
\textbf{Although \advmark is the only viable attack setting in the physical domain, model accuracy is much lower in the digital domain.} We exclude the $\bubblesGray$ dataset as physical attacks against these models were ineffective. The $0/255$ column denotes the  clean accuracy on 500 non-vote bubbles before printing and after the print and scan process.}
\label{tab:digital success under dlr}
\label{tab:physical success dlr}
\end{table*}

\myparab{Correction}
We used \href{https://www.argyllcms.com/#:~:text=ArgyllCMS%20is%20an%20ICC%20compatible,displays%20and%20RGB%20%26%20CMYK%20printers.}{Argyll} with color calibration sheets (e.g., \href{https://www.silverfast.com/products-overview-products-company-lasersoft-imaging/it8-targets-for-scanner-calibration-profiling-for-predictable-brilliant-colors/}{IT8}) to tune the scanner with an \href{https://en.wikipedia.org/wiki/ICC_profile}{ICC profile} for color tonality errors.

\myparab{The drop in accuracy}
Printing and scanning, even with ICC correction techniques in place, introduces additional challenges. Images are visibly different from their digital source. LaserJet printers simulate gray by using dithering patterns to trick the human eye into seeing gray. The net result is that the printed (and re-scanned) bubbles are darker and noisier than the original material the models were trained on. 
As a result, the models classification accuracy drop from over $99\%$ clean accuracy without printing to under  $90\%$, see Table~\ref{table:CleanAccPrinting}. To mitigate, we use a denoising autoencoder that we describe in Appendix~\ref{ssec:denoiser}. Prepending this denoiser to our models makes them accurate at classifying images both before and after printing using the Laserjet printer.

\subsection{Physical Attack Results} 

We run \apgd with DLR loss for $\epsilon$ from $4/255$ to $255/255$ on all of our models.  The examples are fed through the physical extraction pipeline, the denoiser, then the classifier. We consider 500 non-mark bubbles pre-print and post-print. Our attacks only use the classifier weights. Future attacks could incorporate the denoiser weights as part of the backpropagation step.

Results are shown in Table~\ref{tab:physical success dlr}.
For $\epsilon=16/255$ which we judge to be imperceptible in the physical domain, SVM and CaiT are very susceptible to post-print \advmark examples. ResNet-20 and Twins have some resilience but demonstrate robust accuracy less than $1.000$. We note that ResNet-20 and Twins do not demonstrate monotonically decreasing robustness as $\epsilon$ increases. 
This non-monotonic behavior is unexpected, but our attacks are  ``unaware'' of the print-scan noise. Interestingly, digital model robustness appears not to impact physical world resilience.  The highly vulnerable digital models of VGG-16 and Twins are not especially vulnerable to our physical attacks.  However, SVM is vulnerable in both domains.

\myparab{Summary} Adversarial ML involving the physical world is more complicated with several sources of noise that can destroy adversarial signal such as dithering, scanning misalignment, and changes in intensity ranges. Nonetheless for the SVM, ResNet-20, Twins and CaiT $\advmark$ attacks are viable.  As a reminder, a printer can reuse a single attack image.

\section{Conclusion}
\label{sec:conclusion}

Prior work showed that BMDs can alter results and that not all voters check the printed ballot before  tabulation~\cite{bernhard2020can,bernhard2019unclearballot}.  We show that a malicious or compromised ballot printing vendor can print ballots that will be misclassified, even if a voter inspects their ballots. 
While this work focuses on the $l_\infty$ norm for adversarial samples, future work should consider other norms.

Since voting is not an area where one can sacrifice clean accuracy, four main recommendations emerge:
\begin{enumerate}
\itemsep0em
 \item Transparency from vendors on mark interpretations, 
 \item Formalization of voter intent guidelines to allow for standardized labeled datasets with marginal marks, 
 \item  Building mark models that mimic human perception, and 
 \item Explore user interfaces to privately alert voters whenever their ballot features marginal marks and catch risks of misinterpretation.
\end{enumerate}
In summary, our work makes the following contributions: 
\begin{enumerate}
\item We provide four new labeled ballot datasets to the security community.
\item We demonstrate that models trained on ballot data suffer from zero gradients, making standard white-box attacks unusable. We analyze this issue thoroughly and show the DLR loss allows \apgd to be effective.
\item We show that for some models in the physical domain, scanning and printing adversarial examples yields an attack success rate high enough to flip the outcomes of elections with small margins where many voters leave the ballot blank.    
\end{enumerate}

\section*{Acknowledgments} We thank the reviewers and Alexander Russell for their detailed comments and suggestions for improving this manuscript.  This project was jointly funded by grants from the Connecticut Secretary of State's Office and Department of Homeland Security on Grant Number \#DHS-UNO 44-0108-1001-502. L. Michel was partially supported through the Synchrony Endowed Chair. B. Fuller was additional supported through NSF grants \#2141033 and 2232813.

\bibliographystyle{unsrt}
\balance
\bibliography{Voting}
\appendix
\ifnum\conference=0
\section{Full Description of Training Models}
\else
\section{Description of Denoiser Architecture}
\fi
\label{app:full description}

\ifnum\conference=0
We discuss how our classifiers are trained. 

\myparab{Adjusting Training Loader} We have four datasets, Bubbles-Grayscale, Bubbles-RGB, Combined-Grayscale and Combined-RGB.  All models are trained on grayscale with SVM, SimpleCNN, ResNet-20, and Twins also trained on RGB. We use an 80-20 training-validation split. All of these datasets have a disproportional high amount of non-mark examples compared to mark examples. We balance our training loader to have an equal amount of mark and non-mark examples. 


\myparab{Training SVM} SVMs are linear classifier which maximize the distance between the decision boundary and each class in the latent space. They are much simpler than other models. We trained linear SVMs, meaning the linear, or ``identity'', kernel was used. For grayscale datasets, examples were flattened from image size 1 $\times$ 40 $\times$ 50 to a 2000 element tensor as the SVM's input. Likewise, for color datasets, examples were flattened from 3 $\times$ 40 $\times$ 50 to 6000. To match these input dimensions, grayscale SVMs consisted of a single fully-connected layer with 2000 input neurons to a single output neuron. Color SVMs had 6000 input neurons. 

All SVM models were trained using the LinearSVC module from sklearn~\cite{scikit-learn} with the same hyperparameters. We use a linear kernel with $\ell_2$ penalty and balanced class weight. We use a regularization parameter of $1e-8$, a tolerance of $1e-7$, and an intercept scaling factor of 1000. We use primal instead of dual optimization and set the random state to 0. 

\myparab{Training CNNs} Training hyperparameters for the SimpleCNN, VGG-16, and ResNet-20 are listed in Table~\ref{table:training hyperparameters}. Models are trained using CE loss and an Adam optimizer. VGG-16 was not trained on the color datasets.

Table~\ref{table:SimpleCNNArc} outlines the architecture layer-by-layer for the shallow convolutional neural network SimpleCNN. The entire model consists of three convolutional layers that send a one or three channel input (for grayscale and RGB respectively) to 32, to 48, then to 32 channels. The SimpleCNN use in-place ReLU operations with convolution, max pooling, or transposed convolution layer have a padding of 0. (We do the same on the denoiser.) 

\myparab{Training CaiT} We train the \bubblesGray and \combinedGray datasets on CaiT from scratch. The same set of hyperparameters was used for training each dataset for 20 epochs each.  Other hyperparameters include: patch size of 5, embedding dimension of 512, 16 transformer layers, 8 attention heads, MLP dimension of 2048, and dropout and embedding dropout rates of 0.1, along with a layer dropout of 0.05. The models are trained using CE loss and Adam optimizer with learning rate of 5e-5 and batch size of 128.

\begin{table}[t]
\ifnum\conference=1
\tiny
\fi
    \centering
\begin{tabular}{l | l|r r r r r|}
                        
\multirow{2}{*}{Dataset}&\multirow{2}{*}{Model}  &  \multicolumn{5}{c|}{Hyperparameters}\\\cline{3-7}

&& LR & DR & Epochs & WD & BS\\\hline

\multirow{2}{*}{\bubblesGray} & SimpleCNN & 0.01 &  0.9 & 20 & 0.0 & 512\\\cline{3-7}
 & VGG-16 & 0.001 & - & 20 & - & 64\\\cline{3-7}
 & ResNet-20 & 0.1 & 0.9 & 20 & 0.0 & 64\\\cline{1-7}

 \multirow{2}{*}{\combinedGray} & SimpleCNN & 0.001 &  0.9 & 100 & 0.0 & 256\\\cline{3-7}
  & VGG-16 & 0.001 & - & 20 & - & 64\\\cline{3-7}
 & ResNet-20 & 0.0005 & 0.5 & 300 & 0.0 & 128\\\cline{1-7}

 \multirow{2}{*}{\bubblesRGB} & SimpleCNN & 0.01 &  0.9 & 20 & 0.0 & 512\\\cline{3-7}
 & ResNet-20 & 0.1 & 0.9 & 20 & 0.0 & 64\\\cline{1-7}

 \multirow{2}{*}{\combinedRGB} & SimpleCNN & 0.001 &  0.9 & 100 & 0.0 & 512\\\cline{3-7}
 & ResNet-20 & 0.0005 & 0.5 & 300 & 0.0 & 128\\\cline{1-7}
\end{tabular}
\caption{Training hyperparameters for convolutional models SimpleCNN and ResNet-20 across all four datasets. Abbreviations stand for Learning Rate, Dropout Rate, Epochs, Weight Decay, and Batch Size respectively.}
\label{table:training hyperparameters}
\end{table}


\myparab{Training Twins} The Twins-B model was originally pre-trained on ImageNet with RGB images of size 224 $\times$ 224. We leveraged this pre-training to more quickly fine-tune the models trained on the \combinedRGB and \bubblesRGB datasets. Transfer learning is not possible for the \combinedGray and \bubblesGray datasets, however, as grayscale images only have one channel. Thus, these models were trained from scratch. 

Twins was trained with the same set of hyperparameters for each dataset. Each Twins model was trained for 300 epochs with a batch size of 64. All images were upscaled to a resolution of 224 $\times$ 224 during training, evaluation, and attacks. We use an AdamW optimizer with an epsilon of $1e-8$, no beta term, gradient norm clip of 5, a momentum term of 0.9, and a weight decay of 0.05. Dropout was not used during training or evaluation.

\myparab{Additional Adjustments} We list some important optimizations used here:

\begin{enumerate}
    \item We trained our self-attention model, the Twins, on a data range of 0-255 whereas our CNNs (SimpleCNN, VGG-16, and ResNet-20) and linear model, the SVM, were trained on a data range of 0-1. This is because the original Twins model, trained on ImageNet, was trained using a 0-255 range. This allowed us to transfer the model's weights from ImageNet to the \combinedRGB and \bubblesRGB datasets. 

    \item Adding a dropout layer before the final fully-connected layers of each convolutional model and using a high dropout rate helped our combined models achieve high validation accuracy. 

    \item Training on bubbles across all models was significantly easier than training on the combined dataset. Combined models required many epochs to reach a certain accuracy threshold, whereas bubble-only models required a few epochs to reach $100\%$ training accuracy.

    \item To perform CE attacks on the SVM, we add a two-node fully connected layer. The first neuron in our new output FC layer has a weight of -1 and a bias of 1, and the second neuron has a weight of 1 and a bias of 0. This ensures that if the output of our first FC layer is $p$, the output of our second FC layer is $[p, 1-p]$. 

    \item Downsizing 224 $\times$ 224 images to 40 $\times$ 50 then upscaling back to 224 $\times$ 224 removes most of the adversarial signal added to the original images. We need to downsize 224 $\times$ 224 images from Twins for printing and scanning. We run attacks on 40 $\times$ 50 images using Twins by performing an upscaling transformation in Twins' forward pass. 
\end{enumerate}


\myparab{Results} Our fine tuning has allowed almost all of our models across all of our datasets to achieve a $>99\%$ clean validation accuracy on the bubbles dataset for all models except the SVM. Similarly, except the SVM, all combined models achieve a combined validation accuracy above $90\%$.

\begin{table}[t]
\centering
\begin{tabular}{l | r |} 
Layer Type & Parameters \\ 
\hline
Input & (Num Channels, 40, 50) \\ 
Convolution + ReLU& 32$\times$3$\times$3, stride 1\\ 
Max Pooling & 2 $\times$ 2, stride 0 \\ 
Convolution + ReLU& 48 $\times$3$\times$3,  stride 1\\ 
Max Pooling & 2 $\times$ 2,  stride 0 \\ 
Convolution + ReLU & 32 $\times$3$\times$3,  stride 1\\ 
Max Pooling & 2 $\times$ 2, stride 0 \\ 
Dropout & 0.5 \\ 
FC + Softmax & 2
\\ \hline
\multirow{2}{*}{Total Parameters} & Grayscale: 28,818 \\ 
& RGB: 29,394 \\
\hline 
\end{tabular}
\caption{SimpleCNN Architecture. Note the grayscale dataset has one input channel whereas the RGB dataset has three. This causes the total number of training parameters to vary.}
\label{table:SimpleCNNArc}
\end{table}

\subsection{Training Denoiser}
\else
We discuss how our denoiser is trained.  Target model details are available in our \href{https://github.com/VoterCenter/Busting-the-Ballot}{Github repository}. The denoiser reconstructs a clean input from a ``corrupted" version~\cite{vincent2010stacked}. Our denoiser task is to map a post-print bubble back to the original digital bubble. The encoder extracts the feature information from the input (printed and scanned image) through three convolution layers to a  latent space. From this latent space, the decoder uses three transposed convolution layers to recreate the original digital image.  Table~\ref{table:DenoiserArc} outlines the architecture for the denoiser. 

\fi
\label{ssec:denoiser}
\begin{table}[t]
\begin{center}
\begin{tabular}{l|r | r | r| r|}
&\multicolumn{2}{c|}{Sec.~\ref{ssec:classifiers} Model}           & \multicolumn{2}{c|}{Post-Denoiser}                \\ 
& Digital  & Physical & Digital & Physical              \\\hline                    
SVM        & 0.937 & 0.806 & 0.913 & 0.924\\ \hline
SimpleCNN  & 1.000 & 0.750 & 1.000 & 1.000 \\ \hline
ResNet-20  & 1.000 & 0.764 & 0.998 & 0.996 \\ \hline
Twins      & 1.000 & 0.855 & 1.000 & 1.000\\ \hline
\end{tabular}
\end{center}
\caption{Accuracy of all models trained on the $\combinedGray$ dataset on 1000 random classwise balanced $\bubblesGray$ examples pre-print and post-print. \textbf{Our models achieve a higher clean accuracy in the post print domain when using a denoiser.}}
\label{table:CleanAccPrinting}
\end{table}
\begin{center}
\begin{table}[t]
\footnotesize
\begin{tabular}{ l | l | r |}
& Layers & Parameters \\ 
\hline
\multirow{6}{*}{Encoder} & Convolution + ReLU & 32 $\times$ 3 $\times$ 3, pad 0, stride 1 \\ 
& Max Pooling & 2 $\times$ 2, stride 0  \\ 
& Convolution + ReLU & 16 $\times$ 3 $\times$ 3, stride 1 \\ 
& Max Pooling & 2 $\times$ 2, stride 0 \\ 
& Convolution + ReLU & 8 $\times$ 3 $\times$ 3, stride 1 \\ 
& Max Pool 3 & 2 $\times$ 2, stride 0 \\ 
\hline
\multirow{4}{*}{Decoder} & Transpose + ReLU & 8 $\times$ 3 $\times$ 3, pad 0, stride 2 \\ 
& Transpose + ReLU & 16 $\times$ 2 $\times$ 2, stride 2 \\
& Transpose + ReLU & 32 $\times$ 2 $\times$ 2, stride 2 \\ 
& Transpose & 1 $\times$ 3 $\times$ 3, stride 0 \\ 
\hline
\end{tabular}
\caption{Denoising Autoencoder Architecture. The images produced have dimension 42 $\times$ 50, the forward function removes the first and last row.}
\vspace{-.2in}
\label{table:DenoiserArc}
\end{table}
\end{center}
\begin{figure}
    \centering
    \includegraphics[width=.35\textwidth]{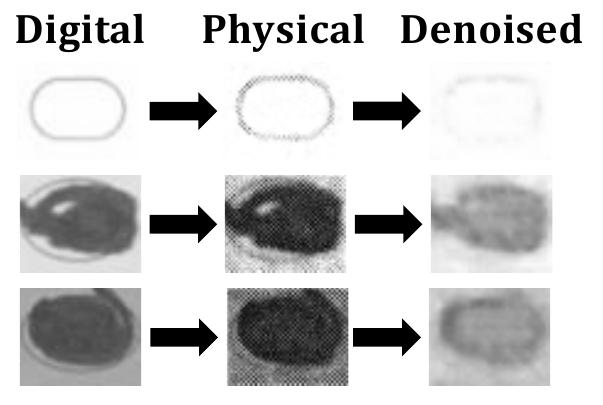}
    \caption[]{Comparison of examples in digital domain, physical domain, and denoiser output. To mitigate the darker tones and dithering in the physical domain, the denoiser adds a blurring effect. }
    \label{fig:denoisedExamples}
\end{figure}
\ifnum\conference=1
\begin{table}[t]
\scriptsize
\else
\begin{table*}[t]
\fi
\begin{tabular}{l | l | l | l | l}
\textbf{Metric} & \textbf{APGD} & \textbf{PGD} & \textbf{MIM} & \textbf{FGSM} \\\hline
\textbf{Dataset} & \bubblesGray / \combinedGray & \bubblesGray / \combinedGray & \bubblesGray / \combinedGray & \bubblesGray / \combinedGray \\\hline
\textbf{Clean Acc.} & 0.9998 / 0.9999 & 0.9998 / 0.9999 & 0.9998 / 0.9999 & 0.9998 / 0.9999 \\\hline
\textbf{4/255} & 1.000 / 0.168 & 1.000 / 0.166 & 1.000 / 0.166 & 1.000 / 0.721 \\\hline
\textbf{8/255} & 1.000 / 0.304 & 1.000 / 0.166 & 1.000 / 0.166 & 1.000 / 0.751 \\\hline
\textbf{16/255} & 1.000 / 0.647 & 1.000 / 0.166 & 1.000 / 0.166 & 0.995 / 0.668 \\\hline
\textbf{32/255} & 1.000 / 0.647 & 1.000 / 0.166 & 1.000 / 0.166 & 0.924 / 0.652 \\\hline
\textbf{64/255} & 1.000 / 0.645 & 1.000 / 0.166 & 1.000 / 0.166 & 0.500 / 0.642 \\\hline
\textbf{255/255} & 1.000 / 0.644 & 1.000 / 0.166 & 1.000 / 0.166 & 0.000 / 0.480 \\
\end{tabular}
    \caption{ResNet-20 Robust Accuracy on \fgsm, \pgd, \mim, and \apgd, using Cross-Entropy loss.}
    \vspace{-.2in}
    \label{table:varyEps resnet}
\ifnum\conference=1
\end{table}
\else
\end{table*}
\fi
We used 100 training epochs with a fixed learning rate of 0.001 and Mean Squared Error loss. To learn the mapping, we train on 10,000 pairs of examples of pre-print/post-print bubbles.  We also include 10,000 digital/digital identity pairs in training to ensure that the denoiser can handle images that have not been printed. No adversarial examples were involved in training.

Results are shown in Table~\ref{table:CleanAccPrinting}, with the denoiser-augmented models demonstrating high $99\%$ accuracy (on three of the models) when given printed or digital examples. SVM pre-print accuracy is lower with an accuracy of $91.3\%$. We print and scan 1000 random classwise balanced \bubblesGray examples. 

Figure~\ref{fig:denoisedExamples} illustrates how the denoiser reverses the darkening and dithering induced by printing.  This carries a cost as images loose some \emph{sharpness}.  Denoiser-prepended models have high accuracy on both digital and physical samples. In the rest of this section, post-print adversarial examples are fed through the denoiser first before classification.

\section{Alternative Attack Results: PGD, MIM, FGSM}
\label{app:other attacks}
For ResNet-20, we believe gradient masking occurs on \apgd~\cite{APGD}, \pgd~\cite{madry2018towards}, \mim~\cite{MIM}, and \fgsm~\cite{FGSMpaper}. Attack results using CE loss are shown in Table~\ref{table:varyEps resnet}. We highlight a few results. First, as previously stated in the literature~\cite{carlini2019evaluating}, unbounded attacks should always produce a robustness of $0\%$. This does \textit{not} happen for any of the attacks with $\epsilon=255/255$ (except FGSM on ResNet-20 Gray-B). \textbf{Every iterative attack (MIM, PGD and APGD) suffers from gradient masking when the CE loss function is used for ResNet-20}. For example, the MIM attack is $16\%$ robust for $\epsilon=255/255$ for ResNet-20 Gray-C when we would expect this number to be $0$. Therefore when working to overcome gradient masking, we focused on refining the APGD attack. We choose to focus on APGD for the following two reasons. First, previous experimental findings~\cite{mahmood2021robustness, APGD} showed that APGD offers superior attack performance in most cases when attacking CNNs and transformers. Second, the literature~\cite{carlini2019evaluating} clearly states using multiple iterative white-box attacks is not a useful form of analysis. Therefore, we did not further develop MIM, PGD and FGSM and instead focused on the APGD attack in our paper. We leave it as an open future work to see whether these attacks could be improved through alternative loss functions. 

\section{Ethical Consideration}
\label{sec:ethics}
This paper addresses the susceptibility of machine learning models trained on voting bubbles to adversarial examples. 
To the best of our knowledge, \textbf{none of the models we have trained and attacked in this paper are used in real tabulating machines}. These models are also not chosen to be similar to models in use by vendors.  Models were chosen to represent different possible design choices.

The marginal marks used in this dataset are designed to explore the boundary of an optical lens scanner and do not result from any election related data.  The empty bubbles are bubbles that were printed by a commercial vendor. They have undergone registration and segmentation using predetermined coordinates.   Marks are on paper printed by the same vendor. There are no obvious indications of the relevant election, ballot type, or what preferences were selected in the processed dataset.

\end{document}